\documentclass[a4paper,11pt]{article}
\usepackage[margin=1in]{geometry}

\usepackage{amsmath}
\usepackage{bm}
\usepackage{natbib}
\usepackage[inline]{enumitem}
\usepackage{amsfonts}
\usepackage{graphicx}
\usepackage{siunitx}
\usepackage{hyperref}
\usepackage{authblk}
\usepackage{pdflscape}
\usepackage{longtable}

\usepackage{ntheorem}
\newtheorem{lemma}{Lemma}
\newtheorem{theorem}{Theorem}

\graphicspath{{Plots/} }

\newcommand{\E}{E}
\newcommand{\pr}{P}

\newcommand{\Var}[1]{\mathrm{var}\left( #1 \right)}
\newcommand{\Varhat}[1]{\widehat{\mathrm{var}}\left( #1 \right)}
\newcommand{\n}[2]{\mathcal{N}\left( #1, #2\right)}

\DeclareMathOperator{\Rank}{Rank}
\DeclareMathOperator{\Dim}{Dim}

\DeclareMathOperator*{\argmin}{arg\,min}
\newcommand{\indep}{\perp\!\!\!\perp}

\newcounter{list1}

\DeclareMathOperator*{\plim}{plim}
\def\T{{ \mathrm{\scriptscriptstyle T} }}

\title{Robust Inference for Mediated Effects in Partially Linear Models}

\author[1]{Oliver Hines}
\author[1,2]{Stijn Vansteelandt}
\author[1]{Karla Diaz-Ordaz}
\affil[1]{Department of Medical Statistics, London School of Hygiene and Tropical Medicine, London, U.K.}
\affil[2]{Department of Applied Mathematics, Computer Science and Statistics, Ghent University, Ghent, Belgium}

\begin{document}

\sisetup{tight-spacing=true}
\maketitle

\begin{abstract}
We consider mediated effects of an exposure, $X$ on an outcome, $Y$, via a mediator, $M$, under no unmeasured confounding assumptions in the setting where models for the conditional expectation of the mediator and outcome are partially linear.

We propose G-estimators for the direct and indirect effect and demonstrate consistent asymptotic normality for indirect effects when models for the conditional means of $M$, or $X$ and $Y$ are correctly specified, and for direct effects, when models for the conditional means of $Y$, or $X$ and $M$ are correct. This marks an improvement, in this particular setting, over previous `triple' robust methods, which do not assume partially linear mean models.

Testing of the no-mediation hypothesis is inherently problematic due to the composite nature of the test (either $X$ has no effect on $M$ or $M$ no effect on $Y$), leading to low power when both effect sizes are small. We use Generalized Methods of Moments (GMM) results to construct a new score testing framework, which includes as special cases the no-mediation and the no-direct-effect hypotheses. The proposed tests rely on an orthogonal estimation strategy for estimating nuisance parameters.

Simulations show that the GMM based tests perform better in terms of power and small sample performance compared with traditional tests in the partially linear setting, with drastic improvement under model misspecification. New methods are illustrated in a mediation analysis of data from the COPERS trial, a randomized trial investigating the effect of a non-pharmacological intervention of patients suffering from chronic pain. An accompanying R package implementing these methods can be found at \texttt{github.com/ohines/plmed}.

\end{abstract}


  \section{Introduction}

  Testing and estimation of mediated effects is important in psychology, sociology, epidemiology, and econometrics, typically as a secondary analysis to understand the mechanism by which an exposure $(X)$ effects an outcome $(Y)$ through a mediating variable $(M)$ \citep{MacKinnon2008,Hayes2018}. When the exposure is binary, one often considers the natural decomposition of the average treatment effect, into a natural indirect effect (NIDE), and natural direct effect (NDE) \citep{Robins1992,Pearl2001}, which, under standard identifiability assumptions (sequential ignorability and consistency), may be written as functionals of the observed distribution \citep{Imai2010a}. These assumptions primarily require observation of a set of variables, $(Z)$, that are sufficient to adjust for confounding of the association between $X$ and $M$ and between $(X,M)$ and $Y$.

  Assuming fully parametric models, maximum likelihood inference for the NIDE and NDE can be based on the so-called mediation formula \citep{Pearl2001,Vanderweele2009,Imai2010a}. Use of this formula gives rise to the popular difference and product-of-coefficient methods \citep{Alwin1975,MacKinnon2002} when simple linear models for the mediator and outcome hold \citep{Vanderweele2009}, but it readily allows extension to non-linear models \citep{Imai2010a}. A key concern about this approach is that misspecification of models for the mediator or outcome can lead to NIDE and NDE estimators with large bias; such misspecification can be difficult to diagnose when some confounders are strongly associated with either the exposure or the mediator \citep{Vansteelandt2012}.

  In contrast, nonparametric inference gives rise to so-called triple robust estimators \citep{Tchetgen2012} of the NIDE and NDE. These are less model-dependent, though still necessitate some form of modelling in view of the curse of dimensionality. In particular, these demand correct specification of an appropriate subset of: (i) the conditional expectation, $\E(Y|M,X,Z)$, (ii) the conditional density of $M$ given $X,Z$, and (iii) the conditional density of $X$ given $Z$. These estimators are called triple robust due to their similarity with `double robust' methods. Double robust methods of the average treatment effect, for example, are Consistent Asymptotically Normal (CAN) provided that either a mean outcome model, or propensity score model is correctly specified \citep{Kang2007}. The triple robust estimator of the marginal NDE and NIDE is CAN provided any pair of (i), (ii), (iii) are correctly specified. These methods are also efficient (under the nonparametric model) provided that (i), (ii), and (iii) are all correctly specified. Additionally, \cite{Tchetgen2014} provide CAN estimators for the parameters indexing a correctly specified parametric model for the conditional NDE given $Z$, provided either (i), or (ii) and (iii) are correctly specified.

  Considering the common use of continuous measurements of mediator and outcome in psychology, which lend themselves to linear modelling, we will consider a different approach in the current work. In particular, we will continue to rely on linear modelling, but in view of the aforementioned concerns about model misspecification, will consider estimation and inference of the NIDE in a semi-parametric partially linear model indexed by $(\beta_1,\beta_2)$ which obeys
  \begin{align}
  \E(M|X,Z)   &=  \beta_1 X + f(Z)  \label{m.mod}  \\
  \E(Y|M,X,Z) &=  \beta_2 M + g(X,Z) \label{y.mod_0}
  \end{align}
  where $g(x,z)$ and $f(x)$ are arbitrary functions. For the NDE, we consider the partially linear model indexed by $(\beta_2,\beta_3)$ which obeys
  \begin{align}
  \E(Y|M,X,Z) &=  \beta_2 M + \beta_3 X + g(Z) \label{y.mod}
  \end{align}
  where $g(z)$ is an arbitrary function. The intersection of these models (i.e. when \eqref{m.mod} and \eqref{y.mod} both hold) is indexed by $\beta=(\beta_1,\beta_2,\beta_3)$. Early work by \cite{Baron1986} on the intersection model defined indirect and direct effects as the product $\beta_1\beta_2$ and coefficient $\beta_3$ respectively, with the total effect given by the sum of the two effects: $\beta_1\beta_2 + \beta_3$. When the exposure is binary and the intersection model holds, then the NIDE, NDE, and average treatment effect reduce to the effect definitions of \cite{Baron1986} (See Section \ref{ident} for details).

  G-estimation is a method of parameter estimation in structural nested models, such as those in \eqref{m.mod} and \eqref{y.mod} developed by James Robins (and collaborators) over a number of years \citep{Robins1994,Vansteelandt2014,Naimi2017}. In the current work, G-estimators for the NIDE and NDE are constructed, assuming that the mediator and outcome mean models are partially linear. Denoting, $h(Z) = \E(X|Z)$, we show that our G-estimator for the NIDE is CAN  when either
  \begin{enumerate}[label=(\alph*)]
  \item \eqref{m.mod} and \eqref{y.mod_0} hold and a parametric model for $f(z)$ is correctly specified
  \item \eqref{m.mod} and \eqref{y.mod} hold and parametric models for both $g(z)$ and $h(z)$ are correctly specified
  \setcounter{list1}{\value{enumi}}
  \end{enumerate}
  and for the NDE, our G-estimator is CAN when either
  \begin{enumerate}[label=(\alph*)]
  \setcounter{enumi}{\value{list1}}
  \item \eqref{y.mod} holds and a parametric model for $g(z)$ is correctly specified
  \item \eqref{m.mod} and \eqref{y.mod} hold and parametric models for both $f(z)$ and $h(z)$ are correctly specified
  \end{enumerate}
  Compared with the triple robust estimators of \cite{Tchetgen2012,Tchetgen2014} (which do not require partial linearity), the proposed G-estimation methods (which do require some partial linearity) have the advantage that conditional densities for $M$ and $X$ do not need to specified, and conditional mean models for $Y$ and $M$ are sufficient to estimate the NIDE and NDE respectively. Extensions to the G-estimation methods where partial linearity is violated are discussed in Section \ref{interactions}.

  We also consider testing of the no-mediation hypothesis ($H_0:\beta_1\beta_2 = 0$) and the no-direct-effect hypothesis ($H_1: \beta_3 = 0$).
  Testing of the no-mediation hypothesis, $H_0$, is problematic since the function used to constrain the hypothesized parameter space ($\psi_0(\beta) = \beta_1\beta_2=0$) has a Jacobian which is full rank almost everywhere, except for a singular point at $\beta_1 = \beta_2 = 0$. This generally gives rise to test statistics with different asymptotic behaviour at this singular point, and in finite samples, tests for $H_0$ which are underpowered in its neighbourhood. We refer the interested reader to \cite{Dufour2013,Drton2016} for further details.

  Ordinary Least Squares (OLS) is routinely used for estimation of the target parameter $\beta$ and nuisance parameter vector $\gamma = (\gamma_{m0},\gamma_m,\gamma_{y0},\gamma_y)$ in the intersection model when $f(z)$ and $g(z)$ are parametrically defined by $f(z) = \gamma_{m0} + \gamma_m^\top z$ and $g(z) = \gamma_{y0} +  \gamma_y^\top z$. Here $\gamma_{m0}$ and $\gamma_{y0}$ represent scalar intercept terms and  $\gamma_m$ and $\gamma_y$ are parameter vectors. Classical tests of the no-mediation hypothesis in this setting are constructed from the squared t-test statistics, $T_j^{(OLS)} = (\hat{\beta}_j^{(OLS)}/\hat{\sigma}_j^{(OLS)})^2$, where for $j=1,2,3$, $\hat{\beta_j}^{(OLS)}$  denotes the OLS estimator, with estimated standard error, $\hat{\sigma}_j^{(OLS)}$.

  Using these squared t-statistics, a Wald test for $H_0$, also known as the Sobel test \citep{Sobel1982} can be constructed, based on the test statistic $W^{(OLS)}$. Alternatively, a joint significance test (also known as a Likelihood Ratio (LR) test)\citep{MacKinnon2002,Giersbergen2014}, has been constructed, based on the statistic $LR^{(OLS)}$. These test statistics are
  \begin{align}
  W^{(OLS)} &=\frac{T_1 T_2}{T_1 + T_2} = \frac{\hat{\beta}_1^2\hat{\beta}_2^2}{\hat{\beta}_1^2\hat{\sigma}_2^2 + \hat{\beta}_2^2\hat{\sigma}_1^2} \label{WALD}\\
  LR^{(OLS)} &= \min(T_1,T_2) \label{LR}
  \end{align}
  where, for readability, the superscript $(OLS)$ has been dropped from all terms on the right hand side. These statistics have received considerable attention, especially due to unexpected properties regarding the relative power of total, direct, and indirect effect tests under different true parameter values \citep{Wang2018,Kenny2014,Fritz2012}.

  We propose two alternative tests based on moment conditions of the G-estimator: a Wald type approach, and an approach analogous to a classical score (or Lagrange Multiplier) test, but derived using a Generalized Methods of Moments (GMM) hypothesis testing framework \citep{Newey1987,Dufour2017}. The relative merits of the new tests against the OLS based tests above are discussed in Section \ref{sect:discussion}. From a robustness perspective, tests based on OLS estimating equations require that \eqref{m.mod} and \eqref{y.mod} hold and $f(z)$ and $g(z)$ are correctly specified, whereas those based on the G-estimation equations inherit the same robustness to model misspecification as the G-estimator itself, provided that nuisance parameters are estimated orthogonally (in a sense defined in Section \ref{nuisance_param}). A simulation study is carried out in Section \ref{sect:sim} to assess the behaviour of the new robust tests in finite samples, followed by an illustration on clinical data in Section \ref{sect:illustration}. All methods are made available through an R package, which can be found at \verb|github.com/ohines/plmed|.

  \section{Identifiability}
  \label{ident}

  Suppose iid data on $(Y,M,X,Z)$ is collected for $n$ individuals. We assume that there exists a potential outcome variable $Y(x,m)$, which expresses the outcome that would have been observed if the exposure and mediator had taken the values $(x,m)$. Similarly, we assume a potential outcome, $M(x)$, corresponding to the mediator if the exposure had taken the value $x$. We define the expected potential outcome
  \begin{align*}
  \eta(x,x^*,z) = \E[Y(x,M(x^*))|Z=z]
  \end{align*}
  for arbitrary $(x,x^*)$ on the support of $X$. We define the NIDE and NDE, conditional on $Z=z$, respectively by
  \begin{align*}
  \E[Y(x_0,M(x_1)) - Y(x_0,M(x_0))|Z=z] \\
  \E[Y(x_1,M(x_1)) - Y(x_0,M(x_1))|Z=z]
  \end{align*}
  for two pre-specified levels of the exposure, $(x_0,x_1)$. For a binary exposure coded 0 or 1, $(x_0,x_1)=(0,1)$, however our definition also permits continuous exposures. Letting $P(m|x,z)$ denote the probability measure of $M$ conditional on $X=x$ and $Z=z$, then
  \begin{align*}
  \eta(x,x^*,z) = \int \E(Y|X=x,M=m,Z=z)dP(m|x^*,z)
  \end{align*}
  under standard identifiability assumptions \citep{Pearl2001,Imai2010a,Imai2010b,Vanderweele2009}. These assumptions require consistency,
  \begin{align*}
  X=x &\implies M(x)=M \text{ almost surely}\\
  X=x \text{ and } M=m &\implies Y(x,m)=Y \text{ almost surely}
  \end{align*}
  and sequential ignorability, which states that for all $m$ on the support of $M$,
  \begin{align*}
  Y(x,m) &\indep M|X=x^*,Z \\
  (Y(x,m),M(x^*)) &\indep X|Z
  \end{align*}
  Here $A\indep B|C$ denotes independence of $A$ and $B$ conditional on $C$. Under these identifiability assumptions and the partial linearity in \eqref{y.mod_0}, then
  \begin{align*}
  \eta(x,x^*,z) = \beta_2 f(x^*,z) + g(x,z)
  \end{align*}
  where $f(X,Z)=\E(M|X,Z)$. We see, therefore, that one obtains the following two expressions for the conditional NIDE and NDE when \eqref{m.mod} and \eqref{y.mod} hold respectively,
  \begin{align*}
  \eta(x_0,x_1,z) - \eta(x_0,x_0,z) &= \beta_1\beta_2(x_1-x_0) \\
  \eta(x_1,x_1,z) - \eta(x_0,x_1,z) &= \beta_3(x_1-x_0)
  \end{align*}
  for all $z$. It follows that when \eqref{m.mod} and \eqref{y.mod_0} hold then the product of coefficients $\beta_1\beta_2$, represents the conditional NIDE per unit change in $X$, and when \eqref{y.mod} holds then $\beta_3$ represents the conditional NDE per unit change in $X$. Since these effects are constant then the marginal effects are equal to the conditional effects. By way of comparison, \cite{Tchetgen2014} consider estimation of a parameter $\psi$ which indexes parametric models for the NDE conditional on $Z=z$, when the exposure is binary,
  \begin{align*}
  \eta(1,1,z) - \eta(0,1,z) &= \delta(z,\psi)
  \end{align*}
  where $\delta(z,\psi)$ is a known function. Our methods, in effect, consider the case $\delta(z,\psi) = \psi$, i.e. that the NDE is constant in subgroups of $Z$, however, we additionally require that partially linear models hold. These assumptions are relaxed in Section \ref{interactions}.

  \section{The G-estimator for mediation}
  \label{sect:g-est}

  Our objective is to derive direct and indirect effect estimators which are asymptotically linear and hence CAN in the sense that they asymptotically follow normal distributions centred at the true value, with variances of order $n^{-1}$. We refer readers to \cite{Kennedy2015} for an introduction to asymptotically linear estimators and influence function theory in causal inference.

  We will consider the target parameter, $\beta = (\beta_1,\beta_2,\beta_3)$ in the intersection model (i.e. when \eqref{m.mod} and \eqref{y.mod} both hold), and begin by introducing parametric working models for $h(z),f(z),g(z)$ which we denote $h(z;\gamma_x),f(z;\gamma_m),g(z;\gamma_y)$ where $h,f,g$ are known differentiable functions parametrized by the nuisance parameter vector $\gamma = (\gamma_x,\gamma_m,\gamma_y)$. This nuisance parameter and the target parameter itself will be estimated jointly. Specifically, we consider an iterative estimation procedure by which the nuisance parameter estimate is obtained from a previous target parameter estimate using a CAN estimator $\hat{\gamma} = \hat{\gamma}(\hat{\beta})$ which is consistent in the sense described by assumptions A1 to A3 below. The target parameter estimate may then be updated using the updated nuisance parameter estimate.  We assume that each component of the nuisance parameter estimator is consistent when the associated part of the model is correctly specified, that is, denoting the correct parameter values by superscript $0$, we assume,
  \begin{enumerate}[label=A\arabic*.]
  \item If $h(z)$ is correctly specified then $\plim \hat{\gamma}_x(\beta) = \gamma_x^0$ for all $\beta$
  \item If $f(z)$ is correctly specified then $\plim \hat{\gamma}_m(\beta)  = \gamma_m^0$ for all $\beta$ such that $\beta_1 = \beta_1^0$
  \item If $g(z)$ is correctly specified then $\plim \hat{\gamma}_y(\beta)  = \gamma_y^0$ for all $\beta$ such that $(\beta_2,\beta_3) = (\beta_2^0,\beta_3^0)$
  \end{enumerate}
  where assumptions A2 and A3 are only well defined when \eqref{m.mod} and \eqref{y.mod} hold respectively. For the target parameter, we propose estimation based on the product of residuals in the intersection model given by the vector $U(\beta,\gamma)$, which we refer to as the G-moment conditions, with components,
  \begin{align}
  U_1(\beta,\gamma) &= \{X-h(Z;\gamma_x)\} \{M-\beta_1 X - f(Z;\gamma_m)\}  \label{G1}\\
  U_2(\beta,\gamma) &= \{M-\beta_1 X - f(Z;\gamma_m)\}\{Y-\beta_2 M - \beta_3 X - g(Z;\gamma_y)\} \label{G2}\\
  U_3(\beta,\gamma) &= \{X-h(Z;\gamma_x)\}\{Y-\beta_2 M - \beta_3 X - g(Z;\gamma_y)\} \label{G3}
  \end{align}
  When all models are correctly specified, these residual products are zero in expectation, i.e. $\E\{U(\beta^0,\gamma^0)\}=0$. The G-estimator for $\beta$, denoted by $\hat{\beta}$ is the value which sets the sample average of moment conditions to zero, that is it solves the system of three equations
  \begin{align}
  \E_n[U\{\hat{\beta},\hat{\gamma}\}] = 0 \label{roots}
  \end{align}
  where $\E_n[.] = n^{-1}\sum_{i=1}^n [.]_i$ is the expectation with respect to the empirical distribution of the data. To examine the behaviour of this estimator under model misspecification, we introduce notation for the probability limit of this G-estimator, $\beta^* = \plim \hat{\beta}$, with associated nuisance parameter estimator limit, $\gamma^* = \plim \hat{\gamma}(\beta^*)$. The probability limit of the G-estimator is the solution to
  \begin{align*}
  \E\{U(\beta^*,\gamma^*)\} = 0 
  \end{align*}
  We additionally assume (A4) that the 3 by 3 matrix,
  \begin{align*}
  \E\left( \frac{\partial U (\beta^*,\gamma^*)}{\partial \beta}\right)
  \end{align*}
  is non-singular. Finally we assume (A5) that $\beta^*$ is unique. This assumption can be partly justified by the linearity of the G-moment conditions, which implies that $\beta^*$ is unique when we disregard the dependence of $\gamma^*$ on $\beta^*$, treating $\gamma^*$ as constant. It is sufficient, therefore, to assume that there is no pathological way by which the nuisance parameter estimator might introduce extra solutions.

  Under assumptions A1 to A5 one can derive the conditions for which the G-estimators of $(\beta_1,\beta_2)$ and of $(\beta_2,\beta_3)$ are consistent, by examining the conditions under which all three moment conditions are zero in expectation. These results are given in Lemmas \ref{NIDE_prop_1} and \ref{NDE_prop_1} respectively. For completeness, the intersection of these two cases, under which $(\beta_1,\beta_2,\beta_3)$ is consistent, is given by Lemma \ref{g_prop}. See supplementary material for proofs.

  \begin{lemma}[Consistency of the G-estimator of $(\beta_1,\beta_2)$]
  \label{NIDE_prop_1}
  Provided the models for $M$ and $Y$ are partially linear in $X$ and $M$ respectively, such that \eqref{m.mod} and \eqref{y.mod_0} both hold, and either
  \begin{enumerate}[label=(\roman*)]
  \item The model for $f(Z)$ is correctly specified
  \item $g(X,Z) = \beta_3 X + g(Z)$ and the models for $g(Z)$ and $h(Z)$ are both correctly specified
  \end{enumerate}
  then $\beta^* = (\beta_1^0,\beta_2^0,\beta_3^*)$, hence the G-estimator is consistent for $(\beta_1,\beta_2)$.
  \end{lemma}

  \begin{lemma}[Consistency of the G-estimator of $(\beta_2,\beta_3)$]
  \label{NDE_prop_1}
  Provided the model for $Y$ is partially linear in $(M,X)$ such that \eqref{y.mod} holds and either
  \begin{enumerate}[label=(\roman*)]
  \item The model for $g(Z)$ is correctly specified
  \item $f(X,Z) = \beta_1 X + f(Z)$ and the models for $f(Z)$ and $h(Z)$ are both correctly specified
  \end{enumerate}
  then $\beta^* = (\beta_1^*,\beta_2^0,\beta_3^0)$, hence the G-estimator is consistent for $(\beta_2,\beta_3)$.
  \end{lemma}

  \begin{lemma}[Robustness of G moment conditions]
  \label{g_prop}
  Provided that the models for $M$ and $Y$ are partially linear in $X$ and $(M,X)$ respectively, such that \eqref{m.mod} and \eqref{y.mod} both hold
  and any pair of
  \begin{enumerate}[label=(\roman*)]
  \item The model for $h(Z)$
  \item The model for $f(Z)$
  \item The model for $g(Z)$
  \end{enumerate}
  are correctly specified, then $\beta^* = (\beta_1^0,\beta_2^0,\beta_3^0)$, hence the G-estimator is consistent for the full target parameter. For proof observe that the conditions in Lemmas \ref{NIDE_prop_1} and \ref{NDE_prop_1} are satisfied.
  \end{lemma}


  Theorem \ref{double} describes the conditions under which the G-estimator will be asymptotically linear.

  \begin{theorem}[Asymptotically linearity of $\hat{\beta}$]
  \label{double}
  Let $\beta^*$ denote the probability limit of the G-estimator, as set out in Lemmas \ref{NIDE_prop_1} to \ref{g_prop}, and assume the nuisance parameter estimator, $\hat{\gamma}(\beta)$ is CAN and obeys assumptions A1 to A3 such that
  \begin{align}
  \hat{\gamma} - \gamma^* = \E_n\{\phi(\beta^*,\gamma^*)\} + o_p\left(n^{-1/2}\right) \label{nuis_estimator}
  \end{align}
  where $o_p$ denotes stochastic order notation so that $A_n = o_p(r_n^{-1})$ means that $A_n r_n \overset{p}{\to} 0$ and $\overset{p}{\to}$ denotes convergence in probability. Then, subject to regularity conditions, the estimator $\hat{\beta}$ is CAN
  \begin{align*}
  \hat{\beta} - \beta^* = \E_n\{\varphi(\beta^*,\gamma^*)\} +  o_p\left(n^{-1/2}\right)
  \end{align*}
  with influence function $\varphi(.)$ given by
  \begin{align}
  \varphi(\beta^*,\gamma^*) &= \E\left\{- \frac{\partial U (\beta^*,\gamma^*)}{\partial \beta}\right\}^{-1}\left(U(\beta^*,\gamma^*) +  \E\left\{\frac{\partial U (\beta^*,\gamma^*)}{\partial \gamma}\right\}\phi(\beta^*,\gamma^*) \right). \label{beta_IF}
  \end{align}
  See Supplementary material for proof.
  \end{theorem}

  The G-estimator of the NIDE under the model in \eqref{m.mod}--\eqref{y.mod_0} is the product $\hat{\beta}_1\hat{\beta}_2$ with influence function,
  \begin{align}
  \omega(\beta^*,\gamma^*) &= \beta_1 \varphi_2(\beta^*,\gamma^*) + \beta_2 \varphi_1(\beta^*,\gamma^*) \label{eta_infl}
  \end{align}
  where $\beta^*= (\beta_1,\beta_2,\beta_3^*)$ and for $j=1,2,3$, $\varphi_j(\beta^*,\gamma^*)$ is the $j$th component of the influence function in \eqref{beta_IF}. Derivation of this influence function can be found in the Supplementary material. Similarly, the G-estimator of the NDE under model \eqref{y.mod} is $\hat{\beta}_3$ with influence function $\varphi_3(\beta^*,\gamma^*)$ where $\beta^*= (\beta_1^*,\beta_2,\beta_3)$ and with consistency guaranteed under the conditions in Lemma \ref{NDE_prop_1}.

  When, in truth, $(\beta_1,\beta_2) = (0,0)$ then the (first-order) influence function in \eqref{eta_infl} is exactly zero. In this case the NIDE estimator $\hat{\beta}_1\hat{\beta}_2$ is asymptotically linear and CAN in the sense that $n^{1/2} \hat{\beta}_1\hat{\beta}_2$ asymptotically follows a normal distribution with zero variance. Multiplying $\hat{\beta}_1\hat{\beta}_2$ by higher powers of $n$ yields more interesting behaviour. When all models are correctly specified, $n\hat{\beta}_1\hat{\beta}_2$ asymptotically follows a `product normal' distribution (the distribution of two mean zero normal variables with known variances) \citep{Aroian1947}.

  \section{Nuisance parameter estimation}
  \label{nuisance_param}

  Theoretical results show that the choice of nuisance parameter estimators does not impact the asymptotic variance of double robust estimators when both working models are correctly specified \citep{Tsiatis2006}. Similarly, in our case, it is straightforward to show that, when the models for $X,M$ and $Y$ are correctly specified then
  \begin{align}
  \E\left\{ \frac{\partial U(\beta^*,\gamma^*)}{\partial \gamma} \right\}  &= 0 \label{Neyman_orthog}
  \end{align}
  So the influence function in \eqref{beta_IF} does not depend on the nuisance influence function $\phi(\beta^*,\gamma^*)$. This property is sometimes referred to as (Neyman) orthogonality \citep{Neyman1959,Chernozhukov2017}, with the intuition that the G-moment conditions are locally insensitive to nuisance parameters when all models are correctly specified. Orthogonal estimators are particularly useful for the construction of score tests, which we describe in Section \ref{score_section}. Moreover, they ensure that our asymptotic results continue to be valid when consistent variable selection procedures (e.g. lasso) are employed for selecting confounders in each of the three working models.

  When \eqref{Neyman_orthog} is not satisfied, as may happen under model misspecification, then the influence function in \eqref{beta_IF} does depend on the nuisance influence function $\phi(\beta^*,\gamma^*)$. Under model misspecification, \eqref{beta_IF} therefore represents a class of G-estimators, indexed by the choice of nuisance parameter estimation method. Choosing the nuisance parameter estimator under misspecification is non-trivial as it may greatly affect the asymptotic variance of the estimator. Various proposals have been suggested for conventional double robust estimators which aim to minimize either the variance under misspecification \citep{Rotnitzky2014} or the bias when models are misspecified \citep{Vermeulen2015,Avagyan2017}. This second approach, referred to as a bias-reduction strategy, involves constructing a nuisance parameter estimator, which is pseudo-orthogonal to the target parameter estimator so that \eqref{Neyman_orthog} is approximately satisfied. In effect, \eqref{Neyman_orthog} is used as a set of moment conditions by which the nuisance parameters are estimated.

  To implement the bias-reduction strategy for our G-estimator, we must first augment the G-moment functions \eqref{G1}--\eqref{G3}, such that each estimating equation has a unique nuisance parameter. In practice, this means, for example, that different estimators of $\gamma_x$ may be used in \eqref{G1} and \eqref{G2}, which are both consistent when $h(z)$ is correctly specified, with the same being true of $\gamma_m$ and $\gamma_y$. Denoting the augmented nuisance parameters with superscript $(1)$ and $(2)$, the augmented moment functions are given by
  \begin{align}
  U_1(\beta,\gamma) &= \left\{X-h\left(Z;\gamma_x^{(1)}\right) \right\} \left\{M-\beta_1 X - f\left(Z;\gamma_m^{(1)}\right)\right\} \\
  U_2(\beta,\gamma) &= \left\{M-\beta_1 X - f\left(Z;\gamma_m^{(2)}\right)\right\} \left\{Y-\beta_2 M - \beta_3 X - g\left(Z;\gamma_y^{(1)}\right)\right\}  \\
  U_2(\beta,\gamma) &= \left\{X-h\left(Z;\gamma_x^{(2)}\right)\right\} \left\{Y-\beta_2 M - \beta_3 X - g\left(Z;\gamma_y^{(2)}\right)\right\}
  \end{align}
  with full nuisance parameter, $\gamma = (\gamma_x^{(1)},\gamma_x^{(2)},\gamma_m^{(1)},\gamma_m^{(2)},\gamma_y^{(1)},\gamma_y^{(2)})$. The bias-reduced nuisance parameter estimator is that which solves
  \begin{equation*}
  \E_n\left\{\frac{\partial U (\hat{\beta},\hat{\gamma})}{\partial\gamma}\right\} = 0
  \end{equation*}
  This estimator is asymptotically linear and obeys the consistency assumptions A1 to A3. For identifiability, we require models where $\Dim(\gamma_x) = \Dim(\gamma_m) = \Dim(\gamma_y)$. Such restrictions are not uncommon e.g. \cite{Rotnitzky2012} and may be satisfied by enlarging the working models. The accompanying \verb|plmed| package implements G-estimation methods with bias-reduced parameter estimation in the setting where $f(z,\gamma_m)$ and $g(z,\gamma_y)$ are linear predictors and $h(z,\gamma_x)$ is modelled by a Generalized Linear Model (GLM).

  \section{Hypothesis testing}
  \label{sect:hyp_test}

  We now consider tests of the null hypothesis, $H_{\alpha}:(\alpha - 1)\beta_1\beta_2 + \alpha \beta_3 = 0$, with $\alpha \in [0,1]$ known. This hypothesis includes the no-mediation hypothesis $(\alpha=0)$ and the no-direct effect hypothesis $(\alpha=1)$ as special cases. We begin by constructing a score test, for general $\alpha$, based on the G-moment conditions. Provided the nuisance parameters are orthogonally estimated, the score test is robust to certain model misspecification. Also, in the specific case where $\alpha = 0$ and the true parameter takes the value $(\beta_1^0,\beta_2^0) = (0,0)$, the score test is conservative, in the sense that the Type I error rate is below the nominal test size.

  The score test is compared with Wald tests for the special cases of the no-mediation hypothesis and the no-direct-effect hypothesis. These Wald tests are constructed using the influence function of the G-estimator, and inherit the robustness properties of the G-estimator, without requiring orthogonal nuisance parameter estimation.

  \cite{Sobel1982} proposed a Wald test of the no-mediation hypothesis, based on the OLS moment conditions. Our Wald test for the no-mediation hypothesis is similar enough to Sobel's work that we shall refer to it as the Robust Sobel test (or Robust Wald test).

  \subsection{The Score Test}
  \label{score_section}

  The score test is based on the observation that, since $\E\{U(\beta^*,\gamma^*)\} = 0$,  the classical central limit theorem implies
  \begin{align}
  n^{1/2} \E_n\{ U(\beta^*,\gamma^*)\} \overset{d}{\to} \n{0}{\E\left\{U(\beta^*,\gamma^*)U(\beta^*,\gamma^*)^\T\right\}} \nonumber \\
  n \E_n\{U(\beta^*,\gamma^*)\}^\T \E\{U(\beta^*,\gamma^*)U(\beta^*,\gamma^*)^\T\}^{-1} \E_n\{ U(\beta^*,\gamma^*)\} \overset{d}{\to} \chi^2_3 \label{gmm:clt}
  \end{align}
  where $\overset{d}{\to}$ denotes convergence in distribution (as $n \to \infty$) and $\n{\mu}{\Sigma}$ and $\chi^2_r$ respectively denote a normal distribution with mean $\mu$ and covariance $\Sigma$, and a chi-squared distribution with $r$ degrees of freedom. The left hand side of \eqref{gmm:clt} is similar in form to a GMM estimator, based on the objective function
  \begin{align*}
  M_n(\beta,\gamma,I) = \E_n\{U(\beta,\gamma)\}^\T I ^{-1} \E_n\{ U(\beta,\gamma)\}\quad  \geq 0 \quad  \forall (\beta,\gamma)
  \end{align*}
  where $I$ is a positive semi-definite 3 by 3 matrix. The GMM estimator of $\beta$ is the minimizer $\argmin_\beta M_n(\beta,\hat{\gamma}(\beta),I)$. In our case the GMM estimator is said to be exactly specified since $\Dim(\beta) = \Dim (U(\beta,\gamma))$. In this exactly specified setting, minimization of the GMM objective function is equivalent to solving \eqref{roots} and the estimator is independent of the choice of weighting matrix, $I$.

  The minimization of $M_n(\beta,\hat{\gamma}(\beta),I)$ over a constrained parameter space, however, may be exploited for hypothesis testing, using results by \cite{Newey1987} for the GMM Two-Step estimator,  later extended by \cite{Dufour2017} to the GMM-Continuous Updating Estimator (CUE), both discussed below.

  Work by \cite{Hansen1982} in the over-specified setting, (i.e. when $\Dim(\beta) < \Dim \{U(\beta,\gamma)\}$), showed that the optimal GMM estimator is constructed using weights proportional to the variance matrix, $I \propto \E\{U(\beta^*,\gamma^*)U(\beta^*,\gamma^*)^\T\}$. This is optimal in the sense that the asymptotic covariance matrix of the resulting estimator is as small as possible (in the positive definite sense) among the class of GMM estimators.

  This optimal choice also lends itself to hypothesis testing, as suggested by \eqref{gmm:clt}. In this work we consider minimization of the objective function, $M_n$, under two proposals. The first (Two-Step) proposal first estimates nuisance parameters and the variance matrix $\E\{U(\beta^*,\gamma^*)U(\beta^*,\gamma^*)^\T\}$. Then, using these initial estimates, constrained estimates of $\beta$ are obtained by a subsequent minimization of the GMM estimator. The second, (CUE) proposal allows the estimates of nuisance parameters and the variance matrix $I$ to be updated continuously. Writing,
  \begin{align*}
  \hat{I}_n(\beta,\gamma) &= \E_n\{U(\beta,\gamma)U(\beta,\gamma)^\T\}
  \end{align*}
  then the proposed Two-Step and CUE objective functions are respectively given by
  \begin{align}
  S(\beta) &= n M_n\Big(\beta,\hat{\gamma}(\hat{\beta}),\hat{I}_n(\hat{\beta},\hat{\gamma}(\hat{\beta}))\Big) \label{tstep_obj} \\
  \tilde{S}(\beta) &= n M_n\Big(\beta,\hat{\gamma}(\beta),\hat{I}_n(\beta,\hat{\gamma}(\beta))\Big) \label{cue_obj}
  \end{align}
  where $\hat{\beta}$ is the unconstrained G-estimate of $\beta$. Defining the null parameter space as $B_{\alpha} = \{\beta| (\alpha - 1)\beta_1\beta_2 + \alpha \beta_3 = 0\}$, the Two-step and CUE score type test statistics may be written as
  \begin{align*}
  S_{\alpha} &= \min_{\beta \in B_{\alpha}} S(\beta) \\
  \tilde{S}_{\alpha} &= \min_{\beta \in B_{\alpha}} \tilde{S}(\beta)
  \end{align*}
  In practice, computation of the Two-Step score statistic may be achieved using the method of Lagrange Multipliers to construct estimating equations for the constrained minimization problem. These may then be solved with a Newton-Raphson scheme. Similarly, computation of the CUE score statistic can be achieved using Lagrange Multipliers to construct estimating equations for $\beta$, however the Newton-Raphson procedure should additionally include the nuisance parameter estimating equations. When the bias-reduced nuisance estimation strategy is used, computation by Newton-Raphson requires that $f(.)$, $g(.)$ and $h(.)$ are twice continuously differentiable.

  To derive the asymptotic distributions of $S_{\alpha}$ and $\tilde{S}_{\alpha}$, we consider the general problem of minimizing over some hypothesis set $B_\psi = \{\beta|\psi(\beta) = 0\}$, where $\psi(.)$ is a differentiable function. For example, when $\psi(\beta) = \beta-\beta^*$ then $B_\psi$ represents a single point, and the asymptotic distribution in \eqref{gmm:clt} is recovered. Theorem \ref{dufour_theorem} gives the general result for the asymptotic distribution of the objective functions, which follows by extending results for the test statistic in Section 5.1 of \cite{Dufour2017}, with related work by \cite{Newey1987}. Our result accommodates nuisance parameter estimation and relies on the orthogonality of the nuisance parameter estimator, see supplementary material for details.

  \begin{theorem}[Constrained GMM]
  \label{dufour_theorem}
  Consider a null hypothesis $H_0: \psi(\beta^*) = 0$, where $\psi$ is a vector of dimension $r \in \{1,2,3\}$ and is continuously differentiable in some non-empty, open neighbourhood, $N$, of the true limiting value $\beta^*$. Provided that for all $\beta \in N$
  \begin{align}
  \Rank \left(\frac{\partial \psi(\beta)}{\partial \beta} \right) = r \label{rank:cond}
  \end{align}
  and $\hat{\gamma}$ is estimated orthogonally, in the sense that \eqref{Neyman_orthog} holds, then for $B_\psi = \{\beta | \psi(\beta) = 0\}$,
  \begin{align*}
  \min_{\beta \in B_\psi } S(\beta) &\overset{d}{\to} \chi^2_r \\
  \min_{\beta \in B_\psi } \tilde{S}(\beta)  &\overset{d}{\to} \chi^2_r.
  \end{align*}
  \end{theorem}

  Applying Theorem \ref{dufour_theorem} to the target hypothesis, $H_\alpha$, we see that the rank condition in \eqref{rank:cond} is not necessarily satisfied for the no-mediation hypothesis $(\alpha = 0)$. Letting $\psi_\alpha (\beta) = (\alpha - 1)\beta_1\beta_2 + \alpha \beta_3$ then
  \begin{align*}
  \Rank \left(\frac{\partial \psi_\alpha(\beta)}{\partial \beta} \right) = \Rank
  \begin{pmatrix}
  (\alpha - 1)\beta_2 \\
  (\alpha - 1)\beta_1 \\
  \alpha
  \end{pmatrix}  = \begin{cases}
  0 & \text{for } \alpha = \beta_1 = \beta_2 = 0\\
  1 & \text{otherwise}
  \end{cases}
  \end{align*}
  Therefore for $\alpha \neq 0$ one may apply the result in Theorem \ref{dufour_theorem} directly to construct a test which rejects $H_\alpha$ when $S_\alpha > c$ for some critical value, $c$. This test size has size $1-F_{\chi^2_1}(c)$ where $F_{\chi^2_1}(x)$ is the distribution function of a $\chi^2_1$ variable. One can show that this test is also a valid test of the no-mediation hypothesis, $H_0$. To do so, we define the null parameter space $B_0 = \{\beta|\beta_1\beta_2=0\}$ and let $C_j = \{\beta|\beta_j=0\}$ for $j= 1,2$ so that $B_0 = C_1 \cup C_2$. Hence
  \begin{align*}
  S_{0} &= \min\left\{ \min_{\beta \in C_1} S(\beta), \min_{\beta \in C_2} S(\beta) \right\} \\
  \tilde{S}_{0} &= \min\left\{ \min_{\beta \in C_1} \tilde{S}(\beta), \min_{\beta \in C_2} \tilde{S}(\beta) \right\} .
  \end{align*}
  Under $H_0$ we know that either $\beta_1=0$ or $\beta_2=0$ and since the constraint function $\psi_j(\beta) = \beta_j$ does satisfy the rank condition in \eqref{rank:cond}, one can show, that under $H_0$,
  \begin{align}
  \sup_{\beta^* \in B_0} \pr_{\beta^*} \left( S_0 > x \right) \to 1 - F_{\chi^2_1} (x) \label{LR_result}  \\
  \sup_{\beta^* \in B_0} \pr_{\beta^*} \left( \tilde{S}_0  > x \right) \to 1 - F_{\chi^2_1} (x) \label{score_result}
  \end{align}
  where $\pr_{\beta^*}$ denotes the probability measure with a true limiting parameter value of $\beta^*$ and $\to$ denotes convergence as $n$ tends to infinity. See supplementary material for details.

  \subsection{Wald Tests}

  Using the asymptotic linearity of $\hat{\beta}_1\hat{\beta}_2$ in \eqref{eta_infl} and under the conditions of Lemma \ref{NIDE_prop_1}, one can demonstrate that $n^{1/2}\hat{\beta}_1\hat{\beta}_2$ asymptotically follows a normal distribution when $(\beta_1,\beta_2) \neq (0,0)$. Estimating the variance of $n^{1/2}\hat{\beta}_1\hat{\beta}_2$ by $\E_n\{\omega^2(\hat{\beta},\hat{\gamma}(\hat{\beta}))\}$ one arrives at a Wald test statistic, $W$, for the no-mediation hypothesis: $\beta_1\beta_2 = 0$
  \begin{align*}
  n^{1/2}(\hat{\beta}_1\hat{\beta}_2 - \beta_1\beta_2) &\overset{d}{\to} \n{0}{\E\{\omega^2(\beta_0,\gamma^*)\}} \\
  W &= \frac{n \hat{\beta}_1^2\hat{\beta}_2^2}{\E_n\{\omega^2(\hat{\beta},\hat{\gamma}(\hat{\beta}))\}} \\
  &= \frac{T_1T_2}{T_1 + T_2 + 2\rho\sqrt{T_1T_2}} \\
  &= \frac{\hat{\beta}^2_1\hat{\beta}^2_2}{\hat{\beta}^2_1 \hat{\sigma}^2_2 + \hat{\beta}^2_2 \hat{\sigma}^2_1 + 2 \hat{\beta}_1 \hat{\beta}_2 \Delta}
  \end{align*}
  where for $j=1,2,3$, the squared t-statistic is represented by $T_j=\hat{\beta}_j^2/\hat{\sigma}^2_j$, and $\rho=\Delta/\hat{\sigma}_1\hat{\sigma}_2$ with $\hat{\sigma}^2_j$ and $\Delta$ given by $n^{-1}\E_n\{\varphi^2_j(\hat{\beta},\hat{\gamma})\}$ and $n^{-1} \E_n\{\varphi_1(\hat{\beta},\hat{\gamma})\varphi_2(\hat{\beta},\hat{\gamma})\}$ respectively.

  The distribution of $W$ is problematic since at the true parameter value $(\beta_1,\beta_2) = (0,0)$, then $\Var{n^{1/2}\hat{\beta}_1\hat{\beta}_2} \to 0$ as $n \to \infty$. A characterisation of Wald-type statistics for testing polynomial constraints with singular points is given by \cite{Dufour2013}. For the constraint $\beta_1\beta_2=0$, \cite{Glonek1993} demonstrated that 
  \begin{align*}
  W \overset{d}{\to} \begin{cases}
  \frac{1}{4} \chi^2_1 & \text{for } \beta_1 = \beta_2 = 0 \\
  \chi^2_1 & \text{otherwise }
  \end{cases}
  \end{align*}
  This result suggests that one may reject the no-mediation hypothesis when the Wald statistic exceeds some critical value, $c$, chosen with reference to the $\chi^2_1$ distribution. Such a test will have size $1-F_{\chi^2_1}(c)$ when the null is satisfied but one of $\beta_1$ or $\beta_2$ is non-zero, and will be conservative when $\beta_1=\beta_2 =0$. The fact that $W$ behaves differently at a singular point is known to greatly restrict the power of the Wald test to detect small indirect effects in finite samples \citep{MacKinnon2002}.

  Construction of a Wald based test for the NDE is fairly trivial. Under the conditions of Lemma \ref{NDE_prop_1}, the squared t-statistic, $T_3 \overset{d}{\to} \chi^2_1$ when $\beta_3^0 =0$.

  \subsection{Comparison of methods}
  \label{sect:discussion}

  Revisiting the classical tests for the no-mediation hypothesis, as given in \eqref{WALD} and \eqref{LR} we see that $ 0 \leq W^{(OLS)} \leq LR^{(OLS)}$ with equality as $T_j^{(OLS)}$ approaches infinity for either $j=1$ or $j=2$, which occurs in the asymptotic limit when $\beta_j\neq 0$. In fact, away from the singularity at $\beta_1^0=\beta_2^0=0$, both statistics have the same $\chi^2_1$ asymptotic distribution and (including the singular point) a test which rejects when $W^{(OLS)}>c$ has equal size to that which rejects when $LR^{(OLS)}>c$ for some critical value $c$. Hence, the test based on $LR^{(OLS)}$ is uniformly more powerful \citep{VanGarderen2019}.

  We highlight this comparison between the two classical tests because it gives some intuition as to why the G-estimation score test, which we argue is analogous to $LR^{(OLS)}$, might be more powerful than the G-estimation Wald test, analogous to $W^{(OLS)}$. The analogy is made clearer by rewriting $LR^{(OLS)}$ as a minimization over an objective function.
  \begin{align}
  S^{(OLS)}(\beta) &= \sum_{j=1}^2 \left(\frac{\hat{\beta}^{(OLS)}_j-\beta_j}{\hat{\sigma}^{(OLS)}_j}\right)^2 \label{ols_obj}\\
  LR^{(OLS)} &= \min_{\{\beta|\beta_1\beta_2=0\}} S^{(OLS)}(\beta) \nonumber
  \end{align}
  This objective function resembles a sum of OLS squared t-statistics, minimization of which (under the constraint $\beta_1\beta_2=0$), either sets $(\beta_1,\beta_2)=(0,\hat{\beta}^{(OLS)}_2)$ or  $(\hat{\beta}^{(OLS)}_1,0)$, thus removing the contribution of a single term from the sum. To demonstrate the analogy between $S^{(OLS)}(\beta)$ and our G-estimation score objective functions in \eqref{tstep_obj} and \eqref{cue_obj}, we consider the case where \eqref{m.mod} and \eqref{y.mod} hold and $f(z),g(z)$ and $h(z)$ are correctly specified.

  In this setting, the G-estimating equations are always orthogonal to the nuisance parameter estimates. For illustration we additionally assume that $\Var{Y|M,X,Z} = \Var{Y|X,Z}$, so that the true covariance matrix, $I = \E\{U(\beta,\gamma)U(\beta,\gamma)^\T\}$ is diagonal. Under these assumptions, the squared t-statistics for the null hypothesis $\beta_j = 0$,  are given by
  \begin{align*}
  T_j &= \frac{\hat{\beta}_j^2}{\hat{\sigma}_j^2} = \frac{n\E_n\{\varphi_j(\hat{\beta}_{-j},\hat{\gamma})\}^2}{\E_n\{\varphi^2_j(\hat{\beta},\hat{\gamma})\}} \text{ for j = 1,2,3}
  \end{align*}
  which reduces to
  \begin{align*}
  T_j &= \frac{n\E_n\{U_j(\hat{\beta}_{-j},\hat{\gamma})\}^2}{\E_n\{U^2_j(\hat{\beta},\hat{\gamma})\}} \text{ for j = 1,2}\\
  T_3&= \frac{n\E_n\{U_3(\hat{\beta}_{-3},\hat{\gamma})\}^2}{\E_n\{U^2_3(\hat{\beta},\hat{\gamma})\}+ \hat{\beta}_1^2 \E_n\{\Varhat{X|Z}\}\E_n\{\Varhat{M|X,Z}\}^{-1} \E_n\{U^2_2(\hat{\beta},\hat{\gamma})\}}
  \end{align*}
  where $\hat{\beta}_{-j}$ denotes the G-estimate of $\beta$ with $j$th parameter set to zero and $\Varhat{.}$ denotes conditional variance estimated using the parameter G-estimates. Note that the denominator of $T_3$ contains an additional term due to the non-zero value of $\E\{\partial U_3 (\beta,\gamma)/\partial \beta_2\}$, which happens to be the only non-zero off-diagonal term of the matrix $\E\{\partial U (\beta,\gamma)/\partial \beta\}$.

  Since the covariance matrix, $I$ is diagonal in this setting, the two-step and CUE objective functions may be written as
  \begin{align*}
  S(\beta) &=  \sum_{j=1}^3 \frac{n\E_n\{U_j(\beta,\hat{\gamma})\}^2}{\E_n\{U^2_j(\hat{\beta},\hat{\gamma}) \}} \\
  \tilde{S}(\beta) &= \sum_{j=1}^3 \frac{n\E_n\{U_j(\beta,\hat{\gamma}(\beta))\}^2}{\E_n\{U^2_j(\beta,\hat{\gamma}(\beta)) \}}
  \end{align*}
  As in \eqref{ols_obj}, these score test objective functions resemble sums of squared t-statistics, making the G-estimation score test analogous to one based on $LR^{(OLS)}$. Theorem \ref{dufour_theorem} may consequently be given the interpretation that the minimization procedure under the null `minimizes out' independent $\chi^2_1$ terms from this score test objective function, leaving a sum of independent $\chi^2_1$ terms equal in number to the dimensions of the constraint.

  \section{Simulation Study}
  \label{sect:sim}

  \subsection{Simulation Study for estimation}

  A simulation study was carried out to examine the bias and variance of NIDE and NDE estimators in finite samples and under model misspecification. G-estimation methods (using bias-reduced nuisance parameter estimation) were compared against the triply robust methods of \cite{Tchetgen2012} (using maximum likelihood methods to fit nuisance parameters). Both the G-estimation methods and triply robust methods (referred to as TTS methods) are available in the \verb|plmed| package. The performance of the proposed score and Wald tests was also compared with classical and TTS derived methods for the no-mediation hypothesis ($H_0$) and the no-direct effect hypothesis ($H_1$). Datasets of size $n$ were generated for different $(\beta_1,\beta_2,\beta_3)$ values using several hierarchical data generating processes, the first of which (Process A) was given by
  \begin{align*}
  Z &\sim \n{0}{1} \\
  X &\sim \text{Bernoulli}\left(\text{expit}( Z + s_x Z^2)\right)\\
  M &\sim \n{\beta_1 X + Z + s_m Z^2 }{1} \\
  Y &\sim \n{\beta_2 M + \beta_3 X + Z + s_y Z^2 }{1}
  \end{align*}
  with $s_x,s_m,s_y \in \{0,1\}$ used to indicate model misspecification and where expit is the inverse-logit function. Additional data generating processes (B and C) used the same models for $Z,X,Y$ with the mediator models instead respectively generated by
  \begin{align*}
  M &= \beta_1 X + Z + s_m Z^2 + \epsilon \\
  M &\sim \text{Bernoulli}\left(\text{expit}(\beta_1 X + Z + s_m Z^2)\right)
  \end{align*}
  where $\epsilon$ follows a Student's t-distribution with 5 degrees of freedom. This Student's t-distribution was chosen as a scenario where an investigator using the TTS methods might fail to correctly model the fat tails of the mediator density. For the G-estimation methods, analysis was conducted under the assumed model
  \begin{align*}
  \E(X|Z) &= \text{expit}( \gamma_{x1} Z + \gamma_{x2})\\
  \E(M|X,Z) &= \beta_1 X + \gamma_{m1} Z + \gamma_{m2} \\
  \E(Y|M,X,Z) &= \beta_2 M + \beta_3 X + \gamma_{y1} Z + \gamma_{y2}
  \end{align*}
  whereas, for processes A and B, the TTS methods additionally assumed that the mediator followed a homoscedastic normal distribution (which is true in the case of process A, but not process B). For process C, the TTS methods assumed that
  \begin{align*}
  M &\sim \text{Bernoulli}\left(\text{expit}(\beta_1 X + \gamma_{m1} Z + \gamma_{m2} )\right)
  \end{align*}
  It follows that for processes A and B, the models assumed by the G-methods are correctly specified when the corresponding misspecification indicator $(s_x,s_m,s_y)$ is equal to zero. For process C, however, we see that \eqref{m.mod} is satisfied only when $\beta_1 = 0$, therefore we expect to obtain valid estimation of the NIDE only when $\beta_1=s_x=s_y=0$. For the NDE, however, \eqref{y.mod} is correct, thus we expect the G-estimation methods to obtain valid inference for the NDE when $s_y=0$.

  To investigate the bias and variance properties of both estimators, two parameter vectors were simulated, $\beta = (0,0,0)$ and $\beta = (1,1,1)$ with sample sizes $n=100,500,1000$ and under various levels of misspecification. 1000 dataset replicates were generated for each simulation, and for each dataset the NIDE, NDE were estimated by both methods. The variance of the G-estimator was also estimated based on influence function theory, and also by bootstrap with 1000 resampling iterations. Monte Carlo estimates for the expectation and variance of each estimator were obtained across the 1000 dataset replicates.

\begin{figure}[tbp]
  \centering
  \includegraphics[scale=0.9]{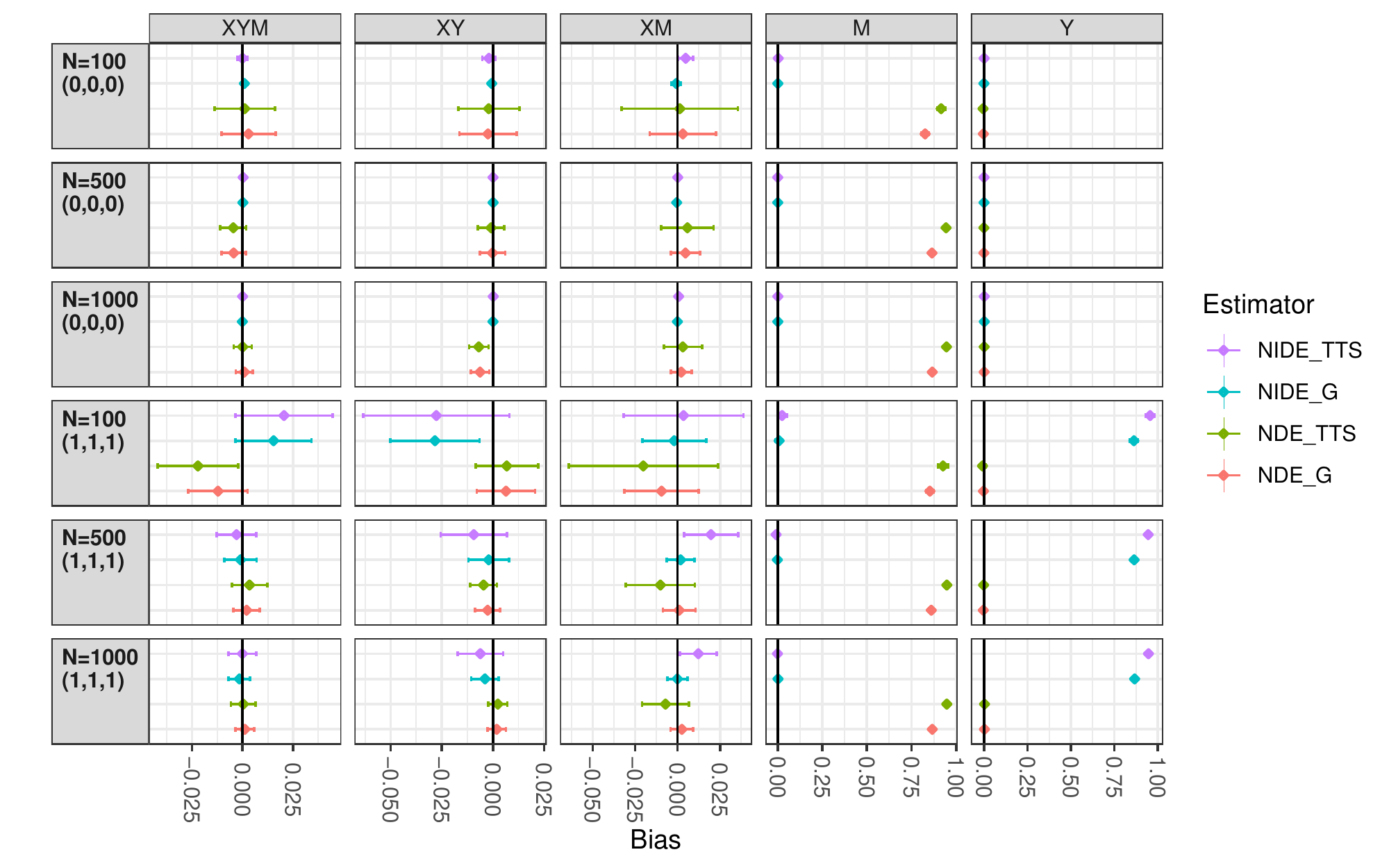}
  \caption{Simulated results to investigate the biases of the NDE and NIDE under data generating process A using G-estimation and TTS methods. The estimated bias from $10^3$ dataset replicates is plotted on the x-axis with error bars giving a 95\% confidence interval of the Monte Carlo estimate. Plots are arranged in a grid where each row represents a
  different sample size and true target parameter value, and the header of each row lists the correctly specified models (those for which the misspecification indicator is equal to zero). We draw the reader's attention to the different scales on the x-axis of these plots}
  \label{sim_plot_bias_het}
  \end{figure}

\begin{figure}[tbp]
  \centering
  \includegraphics[scale=0.9]{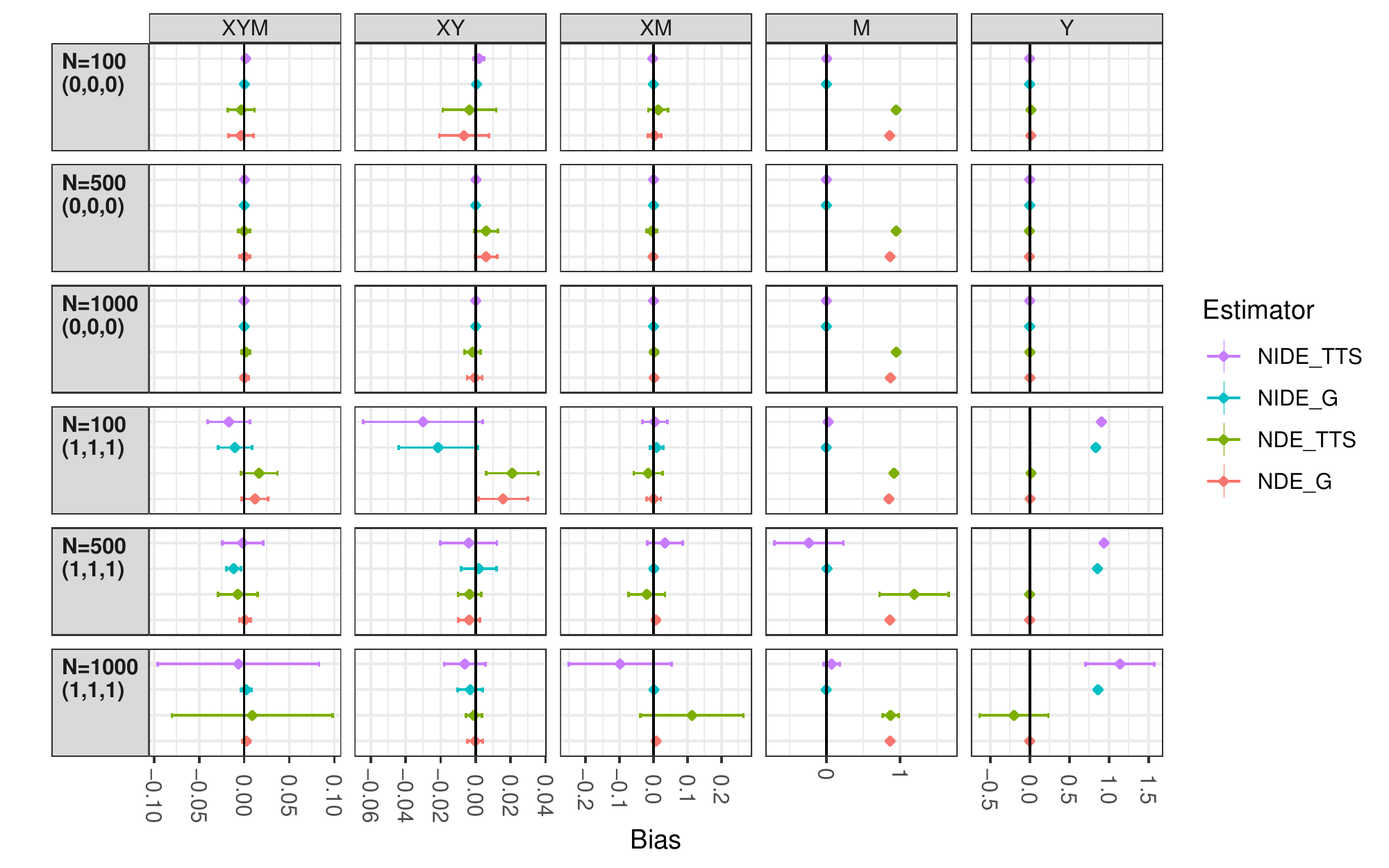}
  \caption{Simulated results to investigate the biases of the NDE and NIDE under data generating process B using G-estimation and TTS methods. The estimated bias from $10^3$ dataset replicates is plotted on the x-axis with error bars giving a 95\% confidence interval of the Monte Carlo estimate. Plots are arranged in a grid where each row represents a
  different sample size and true target parameter value, and the header of each row lists the correctly specified models (those for which the misspecification indicator is equal to zero). We draw the reader's attention to the different scales on the x-axis of these plots}
  \label{sim_plot_bias_t}
  \end{figure}

\begin{figure}[tbp]
  \centering
  \includegraphics[scale=0.9]{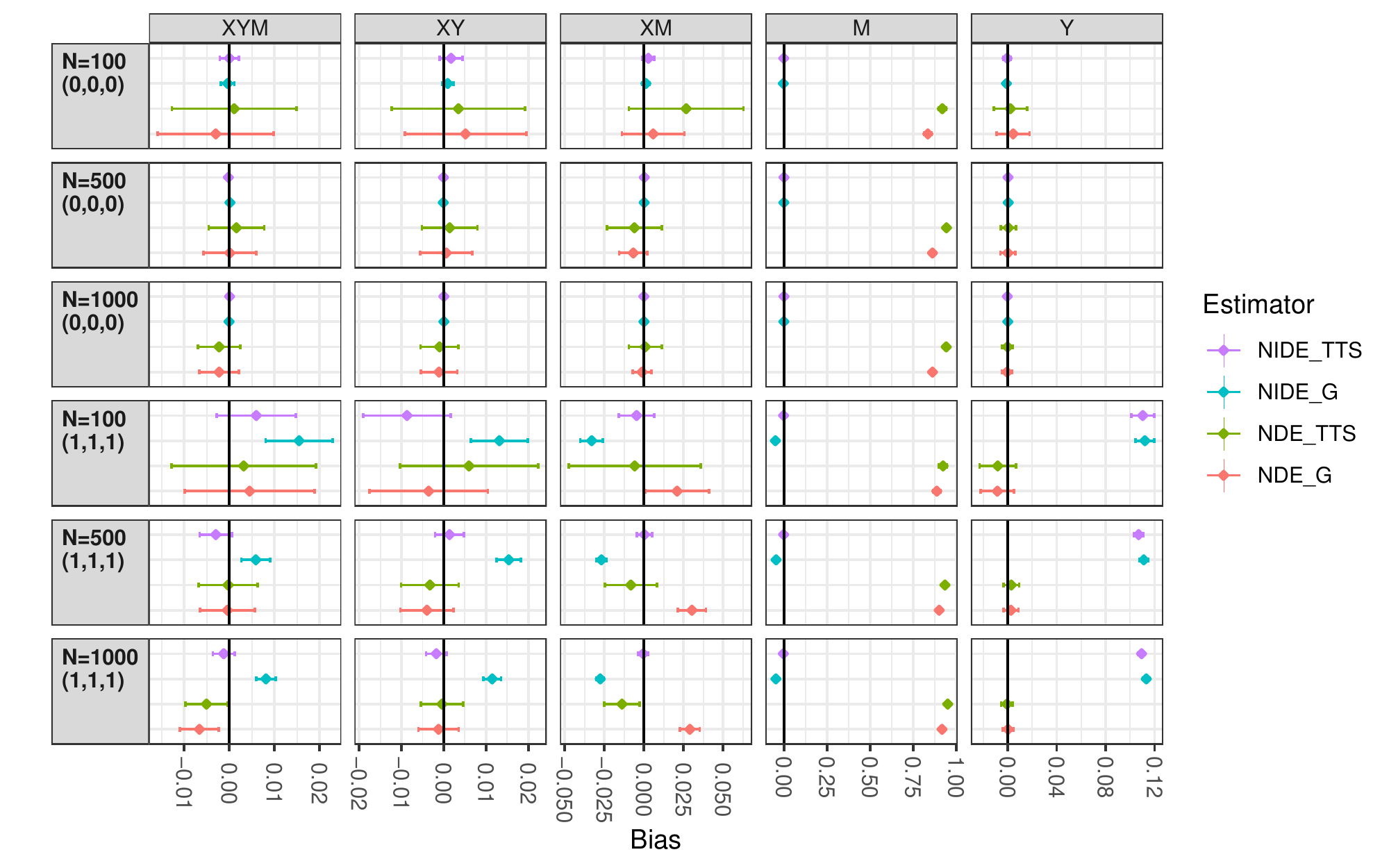}
  \caption{Simulated results to investigate the biases of the NDE and NIDE under data generating process C using G-estimation and TTS methods. The estimated bias from $10^3$ dataset replicates is plotted on the x-axis with error bars giving a 95\% confidence interval of the Monte Carlo estimate. Plots are arranged in a grid where each row represents a
  different sample size and true target parameter value, and the header of each row lists the correctly specified models (those for which the misspecification indicator is equal to zero). We draw the reader's attention to the different scales on the x-axis of these plots}
  \label{sim_plot_bias_log}
  \end{figure}

  Plots of the bias of each estimator can be seen for each data generating process in Figs. \ref{sim_plot_bias_het}, \ref{sim_plot_bias_t}, and \ref{sim_plot_bias_log}. These figures show that when the conditions for the G-estimator are satisfied, the bias remains close to zero, even in small samples. For all data generating process, the standard error in the G-estimator is smaller than that of the TTS methods. This is due to the fact that the G-estimation methods exploit the assumed partial linearity to gain efficiency, whereas the TTS methods do not.

  Interestingly, for process B, the TTS methods perform poorly when the mediator density is misspecified, whereas the G-estimation methods, which do not assume knowledge of the mediator density, perform similarly to data generating process A. This is likely due to large erroneous inverse density weights in the TTS methods. For data generating process C, the G-estimator performs well for the NDE as expected, however, NIDE estimation is biased when $\beta_1 \neq 0$.

  In terms of variance estimation, theoretical results and bootstrap estimation performed similarly with both approximating well the empirical variance of the G-estimator. Full data tables for these simulations can be found in the Supplementary Material.

  \subsection{Simulation Study for hypothesis testing}

\begin{figure}[tbp]
  \centering
  \includegraphics[scale=0.9]{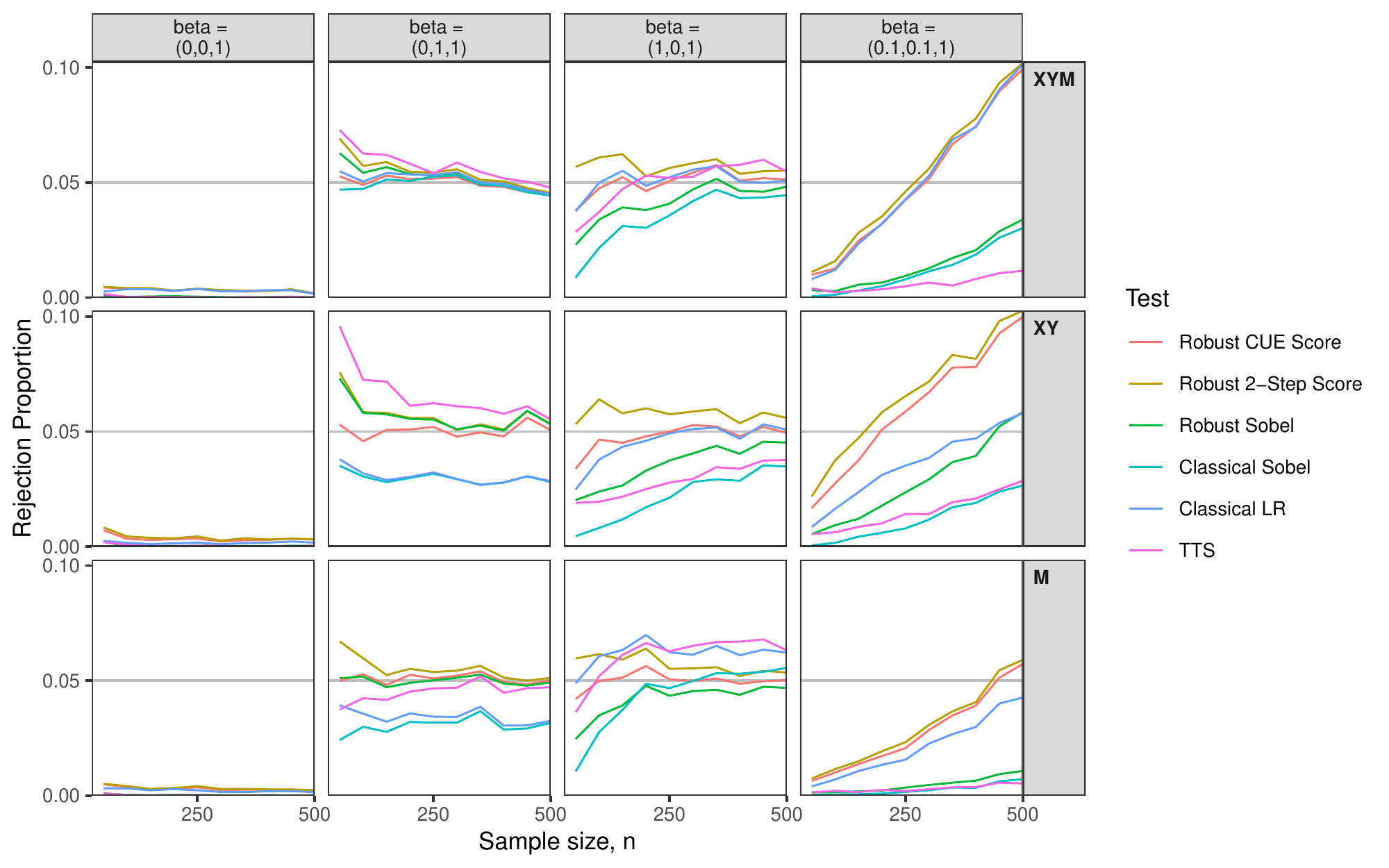}
  \caption{Simulated results from data generating process A of the proportion of the $10^4$ datasets for which the no-mediation hypothesis ($H_0$) is rejected at the 5\% level testing using the CUE score, Two-step score, Robust Sobel, Classical Sobel, Classical LR, and TTS methods. Each column represents a different true $\beta$ parameter, whilst each row gives the models which are correctly specified (those for which the misspecification indicator is equal to zero)}
  \label{sim_plot_1_het}
  \end{figure}

\begin{figure}[tbp]
  \centering
  \includegraphics[scale=0.9]{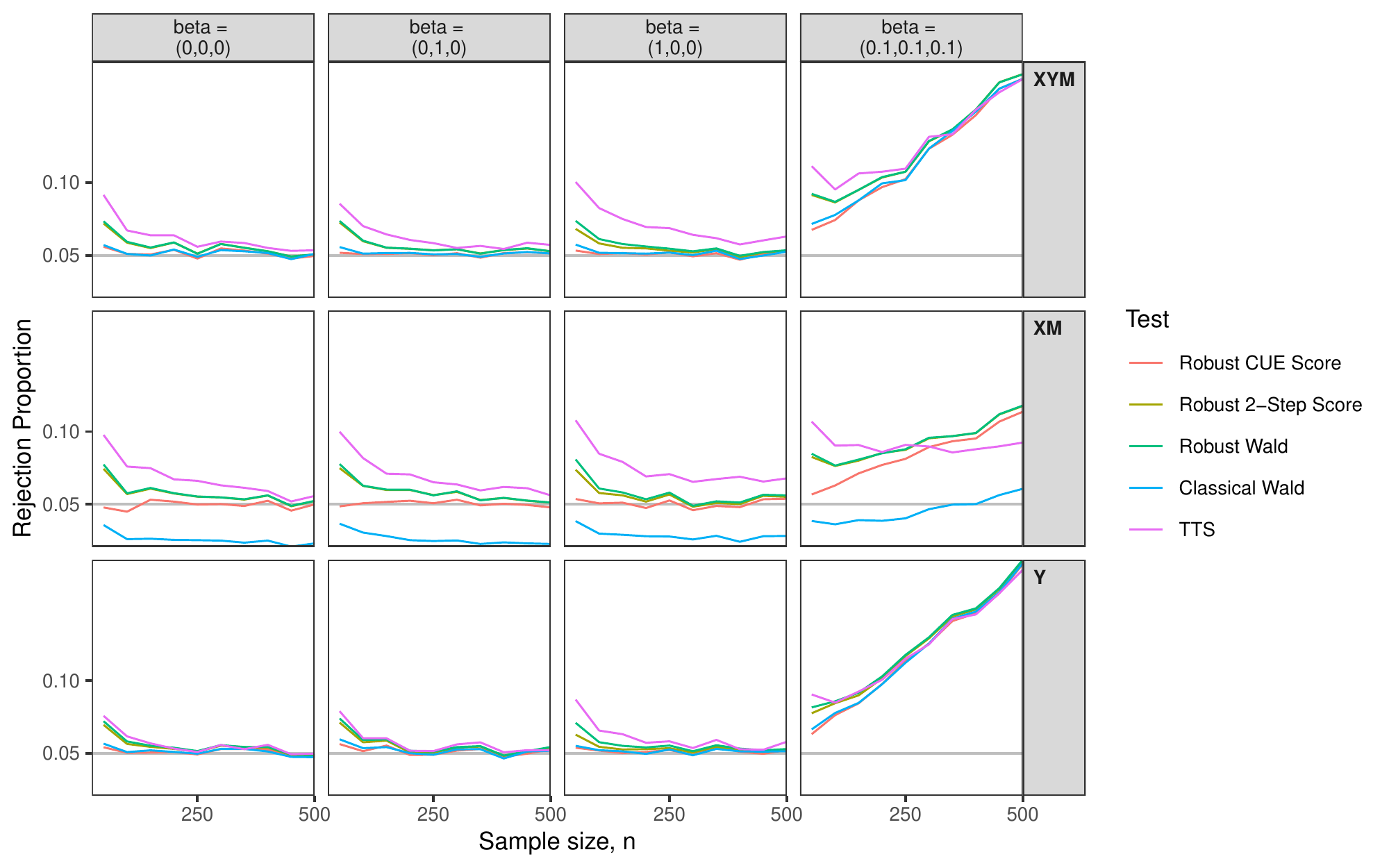}
  \caption{Simulated results from data generating process A of the proportion of the $10^4$ datasets for which the no-direct effect hypothesis ($H_1$) is rejected at the 5\% level testing using the CUE score, Two-step score, Robust Wald, Classical Wald, and TTS methods. Each column represents a different true $\beta$ parameter, whilst each row gives the models which are correctly specified (those for which the misspecification indicator is equal to zero)}
  \label{sim_plot_2_het}
  \end{figure}

  To investigate hypothesis testing methods a greater number of resampled datasets ($10^4$) was used, since the computationally intensive bootstrap variance estimation procedure did not need to be carried out. The proposed tests based on G-estimation were compared with classical (non-robust) methods and Wald tests based on TTS methods. Replicate datasets were generated for $n$ in the range 50 to 500 with various true values of $\beta$ under various levels of misspecification. Figures \ref{sim_plot_1_het} and \ref{sim_plot_2_het} respectively show the proportion of datasets for which the tests of $H_0$ and $H_1$ were rejected at the 5\% level (indicated by a grey line) for data generating process A. When the null is satisfied this rejection proportion corresponds to the Type I error rate, and otherwise corresponds to the statistical power (as in the right most column of both figures). Additional plots for data generating processes B and C can be found in the Supplementary Material, here we consider only data generating process A, as it is representative of all three processes.

  In Fig.\ref{sim_plot_1_het} all testing methods fail to achieve the nominal size when the true parameter takes the value $(\beta_1,\beta_2) = (0,0)$, as expected by theory. All tests, however, do achieve nominal size when one of $\beta_1$ or $\beta_2$ differs from zero, given the requisite misspecification conditions. Under correct specification of all working models, the G-estimation score tests display similar power to the classical $LR$ test, dominating both the robust and classical Sobel tests, which also have similar power to each other. This supports the heuristic argument in Section \ref{sect:discussion}. G-estimation based methods also perform well against those of TTS which, in many cases, seem to converge slowly to the nominal level.

  Although these results suggest that the Two-step procedure has greater power over the CUE score test, the Two-step method appears to have an inflated Type I error rate in small samples, which converges more slowly to the nominal level. This may explain the power discrepancy. This behaviour is reflected also in Fig.\ref{sim_plot_2_het}, where the Robust Wald test suffers from a slightly inflated Type I error rate in small samples. In Fig.\ref{sim_plot_2_het} the robust tests perform better than classical test when $\E(Y|M,X,Z)$ is misspecified, as in the central row.

  \section{Illustrative example: the COPERS trial} \label{Sec:COPERS}
  \label{sect:illustration}

  We now illustrate our G-estimation procedure and hypothesis testing methods, by analysing data from the COPERS  (COping with persistent Pain, Effectiveness Research in Self-management) trial \citep{Taylor2016}. COPERS was a multi-centre, pragmatic, randomized controlled trial examining the effectiveness of a novel non-pharmacological intervention on the management of chronic musculoskeletal pain. Participants in the intervention arm ($n = 384$) were offered to participate in group therapy sessions, while those in the control arm ($n = 300$) received usual care. The group therapy introduced cognitive behavioural approaches to promote self-efficacy in managing chronic pain. The sessions were delivered over three days within the first week with a follow-up session two weeks later.  The control arm participants had no access to the active intervention sessions. Participants and group facilitators were not masked to the study arm they belonged to. The primary outcome, $Y$, was pain-related disability at 12 months, measured on the Chronic Pain Grade (CPG) disability sub-scale. This is a continuous measure on a scale from 0 to 100, with higher scores indicating worse pain-related disability.  The original analysis found no evidence that the COPERS intervention had an effect on improving pain-related disability at 12 months (the average treatment effect on the CPG scale was $-1.0$, with a $95$\% CI of $-4.8$ to $2.7$).

  The COPERS researchers were interested in investigating whether those in the intervention arm that had attended the majority of sessions benefited more from treatment and whether the effect of therapy was mediated by feelings of self-coping with pain. To this effect, trial participants were also asked to fill out the Pain Self-Efficacy Questionnaire (PSEQ)  at 12 weeks (shortly after receiving the intervention), which is intended to measure the participant's confidence to live a normal life despite chronic pain. We will use the score from this questionnaire as a continuous mediator of interest (M).

  Attendance at the 24 group therapy sessions was observed to vary between participants in the intervention arm, with the original investigators considering those who had attended at least 12 sessions as receiving treatment ($A=1,n=260$), with the remaining patients considered as non-treated ($A=0,n=53$). Though planned, a mediation analysis was not performed in the primairy publication \citep{Taylor2016} due to the lack of overall treatment effect. We will examine only the patients randomized to the treatment arm, and conduct two analyses, one in which the exposure of interest, $A$  is binary (with treatment received if attended at least half the sessions), and another where the number of sessions attended defines the continuous exposure (X).

  The baseline covariates included in the original primary analysis are treated, in our analysis, as potential confounders of the three relationships of interest (treatment–mediator, treatment–outcome and mediator–outcome) and thus make up the confounder vector, $Z$, (which also contains an intercept term). These variables are: site of recruitment, employment status, age, gender, Hospital Anxiety and Depression Scale [HADS], Health Education Impact Questionnaire for social integration subscale, and pain-related disability at baseline. We note that these variables may be insufficient to completely adjust for confounding and it is possible that residual unobserved confounding remains. This is an important caveat for the causal interpretation of the mediated effects, however, we proceed under the assumption that residual confounding is negligible.

  Several patients ($n=10$)  were excluded from our analysis as they were missing data on several baseline covariates. Of the remaining patients, some ($n=51$) were missing data on the mediator or outcome variables. It was therefore decided to analyse complete cases ($n=323$), weighting each observation by inverse probability weights derived from a logistic regression model for the missingness probability given $Z$. This method is valid assuming missing-at-random given $Z$. Reported standard errors do not account for the uncertainty in estimating weights, rendering them conservative \citep{Rotnitzky2010}.

  For the binary exposure analysis, a logistic model was assumed for $A$ given $Z$, whilst linear models for the mediator and outcome were assumed (given $(A,Z)$ and $(A,M,Z)$ respectively). i.e.
  \begin{align*}
  \E(A|Z) &= \text{expit}(\gamma_x^\T Z) \\
  \E(M|A,Z) &= \beta_1 A + \gamma_m^\T Z \\
  \E(Y|M,A,Z) &= \beta_2 M + \beta_3 A + \gamma_y^\T Z
  \end{align*}
  The continuous exposure was analysed in a similar fashion, however, using a linear model for $X$ given $Z$, that is
  \begin{align*}
  \E(X|Z) &= \gamma_x^\T Z \\
  \E(M|X,Z) &= \beta_1 X + \gamma_m^\T Z \\
  \E(Y|M,X,Z) &= \beta_2 M + \beta_3 X + \gamma_y^\T Z
  \end{align*}
  Table \ref{copers_tab1} gives mediated effect estimates from the dichotomized exposure analysis by G-estimation, TTS methods (assuming a normally distributed mediator) and OLS. Table \ref{copers_tab2} gives mediated effect estimates from the continuous exposure analysis by G-estimation and OLS. Table \ref{copers_tab3} shows p-values for the no-mediation and no-indirect effect hypotheses from both analyses, obtained using our Robust Sobel and score tests, along with the classical Sobel and LR methods and (for the dichotomized exposure) the TTS methods.

  \begin{table}[htb]
  \caption{Estimated mediation effects for the COPERS trial, treating the exposure as binary and using G-estimation, TTS methods and Ordinary Least Squares}
  \label{copers_tab1}
  \centering
  \begin{tabular}{|r|r|r|r|}
  \hline
  \multicolumn{1}{|c|}{Parameter}  & \multicolumn{1}{c|}{G-estimate(95\% CI)} & \multicolumn{1}{c|}{TTS(95\% CI)} & \multicolumn{1}{c|}{OLS(95\% CI)} \\ \hline
  NDE & 5.31(-4.00,14.6) & -2.96(-15.9,9.97) & 4.83(-2.98,12.6) \\ \hline
  NIDE & -4.60(-7.35,-1.85) & -3.52(-7.47,0.43) & -4.32(-7.05,-1.60) \\ \hline
  $\beta_1$ & 6.06(3.42,8.70) & - & 5.70(2.93,8.47) \\ \hline
  $\beta_2$ & -0.76(-1.07,-0.45) & - & -0.76(-1.06,-0.45)\\ \hline
  \end{tabular}
  \end{table}

  \begin{table}[htb]
  \caption{Estimated mediation effects for the COPERS trial, treating the exposure as continuous and using G-estimation and Ordinary Least Squares}
  \label{copers_tab2}
  \centering
  \begin{tabular}{|r|r|r|r|}
  \hline
  \multicolumn{1}{|c|}{Parameter}  & \multicolumn{1}{c|}{G-estimate(95\% CI)} & \multicolumn{1}{c|}{OLS(95\% CI)} \\ \hline
  NDE & 0.17(-0.23,0.57) & 0.17(-0.18,0.53)  \\ \hline
  NIDE & -0.22(-0.35,-0.09) & -0.22(-0.35,-0.09)  \\ \hline
  $\beta_1$ & 0.29(0.17,0.41) & 0.29(0.17,0.42) \\ \hline
  $\beta_2$ & -0.75(-1.06,-0.44) & -0.75(-1.06,-0.45) \\ \hline
  \end{tabular}
  \end{table}

  \begin{table}[htb]
  \caption{Hypothesis testing results on the COPERS dataset for the null hypotheses $H_\alpha$ ($H_0$ is the no-mediation hypothesis and $H_1$ is the no-direct-effect hypothesis) for the analyses where the exposure is treated as binary and continuous}
  \label{copers_tab3}
  \centering
  \begin{tabular}{|c|l|l|l|}
  \hline
  Null $\alpha$ value & Test   & \multicolumn{1}{l|}{P-Value (binary)} & \multicolumn{1}{l|}{P-Value (continuous)} \\ \hline
  0 & Robust Sobel          & \num{1.18e-3} & \num{9.23e-4} \\ \hline
  0 & Robust score (CUE)    & \num{1.26e-5} & \num{1.28e-5} \\ \hline
  0 & Sobel                 & \num{1.89e-3} & \num{8.50e-4} \\ \hline
  0 & LR                    & \num{5.40e-5} & \num{3.43e-6} \\ \hline
  0 & TTS                   & \num{8.07e-2} & - \\ \hline
  1 & Robust Wald           & 0.264 & 0.396 \\ \hline
  1 & Robust score (CUE)    & 0.256 & 0.397 \\ \hline
  1 & Classical Wald        & 0.225 & 0.341 \\ \hline
  1 & TTS                   & 0.654 & - \\ \hline
  \end{tabular}
  \end{table}

  This mediation analysis sheds some light on the null treatment effect, with significant evidence of an indirect effect. This evidence suggests that session attendance is associated with an increased perception to cope with disability, which in turn, is associated decreased pain-related disability. Interpreting these results causally should be done with caution, due to the possibility of unobserved confounding. Nevertheless, given the possibility of strong mediated effects, researchers interested in cognitive behavioural therapy for chronic pain may want to design add-on interventions that also change self-coping perceptions.

  \section{Extensions}
  \label{interactions}

  Suppose that an investigator is not confident of the partially linear model in \eqref{y.mod_0}, but instead would like to conduct analysis under the semi-parametric model linear model with exposure-mediator interaction,
  \begin{align}
  \E(Y|M,X,Z) = \beta_2 M +\theta XM + g(X,Z) \label{y.mod.int}
  \end{align}
  where $\theta$ is a model parameter, such that when $\theta=0$ there is no exposure-mediator interaction and the model in  \eqref{y.mod_0} is recovered. Under this model, and assuming consistency and sequential ignorability, the potential outcome mean $\eta(x,x^*,z)$ may be written as,
  \begin{align*}
  \eta(x,x^*,z) &= (\beta_2 + \theta x) f(x^*,z) + g(x,z)
  \end{align*}
  As in Section \ref{ident}, one obtains the following two expressions for the conditional NIDE and NDE when \eqref{m.mod} holds and when $g(x,z) = \beta_3 x + g(z)$ respectively,
  \begin{align*}
  \eta(x_0,x_1,z) - \eta(x_0,x_0,z) &= \beta_1(\beta_2 + \theta x_0)(x_1-x_0) \\
  \eta(x_1,x_1,z) - \eta(x_0,x_1,z) &= (\beta_3+\theta f(x_1,z))(x_1-x_0)
  \end{align*}
  The fact that $f(x,z)$ appears in the expression for the NDE gives some indication as to why robust estimation of mediated effects is generally difficult. The solution proposed by \cite{Tchetgen2014} in this setting would be to correctly specify a model for the NDE, $\eta(1,1,z) - \eta(0,1,z)$, implicitly suggesting a correct working model for the conditional expectation of the mediator. This assumption gives the impression of allowing for consistent estimation of the conditional NDE when only the outcome model is correct. The partially linear proposal in the current work is, instead, agnostic to the mediator model, but assumes that $\theta=0$, obtaining valid estimation when the conditional expectation of the outcome is correctly specified (and partially linear in the sense of \eqref{y.mod}).

  For the NIDE, an estimate may be obtained by estimating $(\beta_1,\beta_2,\theta)$, since $(x_1,x_0)$ are known. One might use G-estimation methods to estimate these three parameters and hence the NIDE itself. This might be achieved by estimation of $(\beta_1,\beta_2,\beta_3,\theta)$ in the intersection model using the set of G-estimation moment conditions
  \begin{align*}
  U_1(\beta,\gamma) &= \{X-h(Z;\gamma_x)\} \{M-\beta_1 X - f(Z;\gamma_m)\} \\
  U_2(\beta,\gamma) &= \{M-\beta_1 X - f(Z;\gamma_m)\}\{Y-\beta_2 M - \theta XM - \beta_3 X - g(Z;\gamma_y)\} \\
  U_3(\beta,\gamma) &= \{X-h(Z;\gamma_x)\}\{Y-\beta_2 M - \theta XM - \beta_3 X - g(Z;\gamma_y)\} \\
  U_4(\beta,\gamma) &= X\{M-\beta_1 X - f(Z;\gamma_m)\}\{Y-\beta_2 M - \theta XM - \beta_3 X - g(Z;\gamma_y)\}
  \end{align*}
  Hence, no additional working models are required. Using methods similar to those used to show Lemma \ref{NIDE_prop_1}, one can show that these moment conditions have zero expectation (for some $\beta_3$) when \eqref{m.mod} and \eqref{y.mod.int} hold and either $f(z)$ is correctly specified, or $g(x,z) = \beta_3 x + g(z)$ and both $g(z)$ and $h(z)$ are correctly specified. Results concerning estimation and testing could also be extended to account for the fourth moment conditions.

  In a similar way, additional estimating equations could also be included to estimate parameters associated with counfounder interactions, such as interactions of the form $Z_jX$ in the mediator or outcome model, or of the form $Z_jM$ in the outcome model, where $j$ indexes the set of counfounders, $Z$. Alternatively, when the confounder variable, $Z_j$, is categorical then the partial linearity assumptions, \eqref{m.mod}, \eqref{y.mod_0}  and \eqref{y.mod}, may be satisfied within certain population subgroups, i.e. the target parameters $(\beta_1,\beta_2,\beta_3)$ differ between subgroups. In this setting, one simple strategy is to estimate mediation effects for each subgroup and take a weighted average of these effects. In practice this could be achieved by passing indicator weights to the \verb|plmed| fitting functions.

  Finally, we consider how the proposed G-estimation NIDE estimator (i.e. using moment conditions moment \eqref{G1}--\eqref{G3}) performs when the true data generating distribution follows \eqref{m.mod} and \eqref{y.mod.int}. Lemma \ref{NIDE_prop_1} considers the special case where $\theta = 0$. In general, however, provided $f(z)$ is correctly specified then,
  \begin{align*}
  \beta_1^* \beta_2^* &= \beta_1(\beta_2 + \theta \bar{x}) \\
  \bar{x} &= \frac{\E[X\mathrm{var}(M|X,Z)]}{\E[\mathrm{var}(M|X,Z)]}
  \end{align*}
  See supplementary material for details. For continuous exposures $\beta_1^* \beta_2^*$ may thus be interpreted as the NIDE per unit change in $X$ at $x_0 = \bar{x}$. For binary exposures, however, the potential outcome when $X=\bar{x}$ is not well defined. For an analogous interpretation, one might consider a conditional indirect effect defined by
  \begin{align*}
  \Psi(x) &= \eta(x,1,z) - \eta(x,0,z) \\
  &= \beta_1(\beta_2 + \theta x)
  \end{align*}
  where $x$ is some level of the exposure. For binary exposures, the G-estimator returns a weighted average of $\Psi(x)$, which retains the interpretation of an indirect effect
  \begin{align*}
  \beta_1^* \beta_2^* &= \frac{\E[\Psi(X)\mathrm{var}(M|X,Z)]}{\E[\mathrm{var}(M|X,Z)]}
  \end{align*}
  By comparison, in this setting where the outcome model is misspecified, the TTS methods return $\Psi(1)$, provided that the exposure is binary and that $h(z)$ and the conditional density of $M$ given $X$ and $Z$ are both correctly specified.

  \section{Discussion}

  The main contribution of the current paper is a practical and robust method for carrying out inference of mediated effects in settings where partial linearity of mediator and outcome conditional expectations can be assumed and the vector of variables needed to control for confounding is low dimensional. We recommend estimation of the NIDE and NDE by G-estimation, in settings where partial linearity may be assumed, but a mediator density function may be difficult to estimate, as required by the methods of \cite{Tchetgen2012}. This is for instance the case when analysing continuous mediators, as often encountered in applications in psychology. Compared with OLS, the G-estimators are consistent for the NIDE and NDE, under misspecification of mediator and outcome models and outcome models respectively. The variance of these estimators may be estimated by bootstrap or using asymptotic results with both giving similar results. We also make available the methods in the R package \verb|plmed|, which calculates the NDE and NIDE by G-estimation with variances estimated using asymptotic results.

  In terms of hypothesis testing we recommend the robust CUE score test over Two-step robust score methods, due to its faster convergence of the Type I error rate to the nominal size in small samples, and improved power to detect small NIDEs over Wald based testing methods, as demonstrated in simulation studies.

  In future work one can hope to extend the proposed G-estimation results to the high-dimentional setting, where the number of parameters indexing nuisance models does not need to be small (compared with the number of observations). In particular, by exploiting the orthogonality of the G-estimator when the exposure and outcome models are both correct, one can show that valid inference of the average treatment effect is obtained even when cross-validated data-adaptive methods (e.g. lasso or machine learning) are used to estimate the nuisance models (such as $f(z),g(z),$ and $h(z)$ in the current paper), with the assumption that such methods will converge to the true model at a sufficiently fast rate and an appropriate sample splitting scheme is applied \citep{Chernozhukov2017}.

  Other work by \cite{Dukes2019} obtains valid inference of the average treatment effect using G-estimators when nuisance models are fitted using the bias-reduction strategy with a lasso $l_1$ penalty on the nuisance parameter. This work does not require sample splitting nor does it require that both the exposure and outcome models converge to the truth. Indeed their methods are valid even when the number of confounding variables is allowed to grow with sample size, provided certain sparsity assumptions on the nuisance parameter are satisfied. These methods may be applied directly to the methods in the current paper, with Theorem \ref{dufour_theorem}, holding even when the nuisance estimator is orthogonal and penalized (provided that it continues to be orthogonal as for instance in \cite{Dukes2019}).


\bibliography{refs}

\begin{thebibliography}{}

\bibitem[Alwin and Hauser, 1975]{Alwin1975}
Alwin, D.~F. and Hauser, R.~M. (1975).
\newblock {The Decomposition of Effects in Path Analysis}.
\newblock {\em American Sociological Review}, 40(1):37.

\bibitem[Aroian, 1947]{Aroian1947}
Aroian, L.~A. (1947).
\newblock {The Probability Function of the Product of Two Normally Distributed
  Variables}.
\newblock {\em The Annals of Mathematical Statistics}, 18(2):265--271.

\bibitem[Avagyan and Vansteelandt, 2017]{Avagyan2017}
Avagyan, V. and Vansteelandt, S. (2017).
\newblock {Honest data-adaptive inference for the average treatment effect
  under model misspecification using penalised bias-reduced double-robust
  estimation}.
\newblock arXiv:1708.03787.
\newblock Forthcoming in Biostatistics and Epidemiology.

\bibitem[Baron and Kenny, 1986]{Baron1986}
Baron, R.~M. and Kenny, D.~A. (1986).
\newblock {The moderator mediator variable distinction in social psychological
  research: Conceptual, strategic, and statistical considerations.}
\newblock {\em Journal of Personality and Social Psychology}, 51(6):1173--1182.

\bibitem[Chernozhukov et~al., 2017]{Chernozhukov2017}
Chernozhukov, V., Chetverikov, D., Demirer, M., Duflo, E., Hansen, C., and
  Newey, W. (2017).
\newblock {Double/Debiased/Neyman Machine Learning of Treatment Effects}.
\newblock {\em American Economic Review}, 107(5):261--265.

\bibitem[Drton and Xiao, 2016]{Drton2016}
Drton, M. and Xiao, H. (2016).
\newblock {Wald tests of singular hypotheses}.
\newblock {\em Bernoulli}, 22(1):38--59.

\bibitem[Dufour et~al., 2013]{Dufour2013}
Dufour, J.-M., Renault, E., and Zinde-Walsh, V. (2013).
\newblock {Wald tests when restrictions are locally singular}.
\newblock arXiv:1312.0569.

\bibitem[Dufour et~al., 2017]{Dufour2017}
Dufour, J.~M., Trognon, A., and Tuvaandorj, P. (2017).
\newblock {Invariant tests based on M-estimators, estimating functions, and the
  generalized method of moments}.
\newblock {\em Econometric Reviews}, 36(1-3):182--204.

\bibitem[Dukes and Vansteelandt, 2019]{Dukes2019}
Dukes, O. and Vansteelandt, S. (2019).
\newblock {Uniformly valid confidence intervals for conditional treatment
  effects in misspecified high-dimensional models}.
\newblock arXiv:1903.10199.

\bibitem[Fritz et~al., 2012]{Fritz2012}
Fritz, M.~S., Taylor, A.~B., and MacKinnon, D.~P. (2012).
\newblock {Explanation of Two Anomalous Results in Statistical Mediation
  Analysis}.
\newblock {\em Multivariate Behavioral Research}, 47(1):61--87.

\bibitem[Giersbergen, 2014]{Giersbergen2014}
Giersbergen, N. (2014).
\newblock Inference about the indirect effect: a likelihood approach.
\newblock UvA-Econometrics Working Papers 14-10, Universiteit van Amsterdam,
  Dept. of Econometrics.

\bibitem[Glonek, 1993]{Glonek1993}
Glonek, G. F.~V. (1993).
\newblock {On the Behaviour of Wald Statistics for the Disjunction of Two
  Regular Hypotheses}.
\newblock {\em Journal of the Royal Statistical Society. Series B
  (Methodological)}, 55(3):749--755.

\bibitem[Hansen, 1982]{Hansen1982}
Hansen, L.~P. (1982).
\newblock {Large Sample Properties of Generalized Method of Moments
  Estimators}.
\newblock {\em Econometrica}, 50(4):1029.

\bibitem[Hayes, 2018]{Hayes2018}
Hayes, A.~F. (2018).
\newblock {\em {Introduction to mediation, moderation, and conditional process
  analysis: a regression-based approach}}.
\newblock Guilford Press, New York, NY, 2 edition.

\bibitem[Imai et~al., 2010a]{Imai2010a}
Imai, K., Keele, L., and Tingley, D. (2010a).
\newblock {A General Approach to Causal Mediation Analysis}.
\newblock {\em Psychological Methods}, 15(4):309--334.

\bibitem[Imai et~al., 2010b]{Imai2010b}
Imai, K., Keele, L., and Yamamoto, T. (2010b).
\newblock {Identification, inference and sensitivity analysis for causal
  mediation effects}.
\newblock {\em Statistical Science}, 25(1):51--71.

\bibitem[Kang and Schafer, 2007]{Kang2007}
Kang, J.~D. and Schafer, J.~L. (2007).
\newblock {Demystifying double robustness: A comparison of alternative
  strategies for estimating a population mean from incomplete data}.
\newblock {\em Statistical Science}, 22(4):523--539.

\bibitem[Kennedy, 2015]{Kennedy2015}
Kennedy, E.~H. (2015).
\newblock {Semiparametric theory and empirical processes in causal inference}.
\newblock arXiv:1510.04740.

\bibitem[Kenny and Judd, 2014]{Kenny2014}
Kenny, D.~A. and Judd, C.~M. (2014).
\newblock {Power Anomalies in Testing Mediation}.
\newblock {\em Psychological Science}, 25(2):334--339.

\bibitem[MacKinnon, 2008]{MacKinnon2008}
MacKinnon, D.~P. (2008).
\newblock {\em {Introduction to statistical mediation analysis.}}
\newblock Multivariate applications series. Taylor {\&} Francis Group/Lawrence
  Erlbaum Associates, New York, NY.

\bibitem[MacKinnon et~al., 2002]{MacKinnon2002}
MacKinnon, D.~P., Lockwood, C.~M., Hoffman, J.~M., West, S.~G., and Sheets, V.
  (2002).
\newblock {A comparison of methods to test mediation and other intervening
  variable effects.}
\newblock {\em Psychological Methods}, 7(1):83--104.

\bibitem[Naimi et~al., 2017]{Naimi2017}
Naimi, A.~I., Cole, S.~R., and Kennedy, E.~H. (2017).
\newblock {An introduction to g methods}.
\newblock {\em International journal of epidemiology}, 46(2):756--762.

\bibitem[Newey and West, 1987]{Newey1987}
Newey, W.~K. and West, K.~D. (1987).
\newblock {Hypothesis Testing with Efficient Method of Moments Estimation}.
\newblock {\em International Economic Review}, 28(3):777.

\bibitem[Neyman, 1959]{Neyman1959}
Neyman, J. (1959).
\newblock {Optimal Asymptotic Tests of Composite Statistical Hypotheses}.
\newblock In Grenander, U., editor, {\em Probability and Statistics: The Harald
  Cramer Volume}, pages 213--234. Almqvist and Wiskell, Stockholm.

\bibitem[Pearl, 2001]{Pearl2001}
Pearl, J. (2001).
\newblock {Direct and Indirect Effects}.
\newblock In {\em Proc.\ 17th Conference on Uncertainy in Articial
  Intelligence}, pages 411--420. San Francisco. Morgan Kaufmann.

\bibitem[Robins, 1994]{Robins1994}
Robins, J.~M. (1994).
\newblock {Correcting for non-compliance in randomized trials using structural
  nested mean models}.
\newblock {\em Communications in Statistics - Theory and Methods},
  23(8):2379--2412.

\bibitem[Robins and Greenland, 1992]{Robins1992}
Robins, J.~M. and Greenland, S. (1992).
\newblock {Identifiability and Exchangeability for Direct and Indirect
  Effects}.
\newblock {\em Epidemiology}, 3(2):143--155.

\bibitem[Rotnitzky et~al., 2012]{Rotnitzky2012}
Rotnitzky, A., Lei, Q., Sued, M., and Robins, J.~M. (2012).
\newblock {Improved double-robust estimation in missing data and causal
  inference models}.
\newblock {\em Biometrika}, 99(2):439--456.

\bibitem[Rotnitzky et~al., 2010]{Rotnitzky2010}
Rotnitzky, A., Li, L., and Li, X. (2010).
\newblock {A note on overadjustment in inverse probability weighted
  estimation}.
\newblock {\em Biometrika}, 97(4):997--1001.

\bibitem[Rotnitzky and Vansteelandt, 2014]{Rotnitzky2014}
Rotnitzky, A. and Vansteelandt, S. (2014).
\newblock Double-robust methods.
\newblock In Molenberghs, G., Fitzmaurice, G., Kenward, M., Tsiatis, A., and
  Verbeke, G., editors, {\em Handbook of missing data methodology}, chapter~9,
  pages 185--212. CRC Press, New York.

\bibitem[Sobel, 1982]{Sobel1982}
Sobel, M.~E. (1982).
\newblock {Asymptotic Confidence Intervals for Indirect Effects in Structural
  Equation Models}.
\newblock {\em Sociological Methodology}, 13(1982):290.

\bibitem[Taylor et~al., 2016]{Taylor2016}
Taylor, S.~J., Carnes, D., Homer, K., Pincus, T., Kahan, B.~C., Hounsome, N.,
  Eldridge, S., Spencer, A., Diaz-Ordaz, K., Rahman, A., Mars, T.~S., Foell,
  J., Griffiths, C.~J., and Underwood, M.~R. (2016).
\newblock {Improving the self-management of chronic pain: COping with
  persistent Pain, Effectiveness Research in Self-management (COPERS)}.
\newblock {\em Programme Grants for Applied Research}, 4(14):1--440.

\bibitem[{Tchetgen Tchetgen} and Shpitser, 2012]{Tchetgen2012}
{Tchetgen Tchetgen}, E.~J. and Shpitser, I. (2012).
\newblock {Semiparametric theory for causal mediation analysis: Efficiency
  bounds, multiple robustness and sensitivity analysis}.
\newblock {\em Annals of Statistics}, 40(3):1816--1845.

\bibitem[Tchetgen~Tchetgen and Shpitser, 2014]{Tchetgen2014}
Tchetgen~Tchetgen, E.~J. and Shpitser, I. (2014).
\newblock {Estimation of a semiparametric natural direct effect model
  incorporating baseline covariates}.
\newblock {\em Biometrika}, 101(4):849--864.

\bibitem[Tsiatis, 2006]{Tsiatis2006}
Tsiatis, A.~A. (2006).
\newblock {\em {Semiparametric Theory and Missing Data}}.
\newblock Springer Series in Statistics. Springer New York, New York, NY.

\bibitem[van Garderen and van Giersbergen, 2019]{VanGarderen2019}
van Garderen, K.~J. and van Giersbergen, N. (2019).
\newblock {Almost Similar Tests for Mediation Effects Hypotheses with
  Singularities}.
\newblock arXiv:2012.11342.

\bibitem[VanderWeele and Vansteelandt, 2009]{Vanderweele2009}
VanderWeele, T.~J. and Vansteelandt, S. (2009).
\newblock {Conceptual issues concerning mediation, interventions and
  composition}.
\newblock {\em Statistics and Its Interface}, 2(4):457--468.

\bibitem[Vansteelandt, 2012]{Vansteelandt2012}
Vansteelandt, S. (2012).
\newblock {Understanding counterfactual-based mediation analysis approaches and
  their differences}.
\newblock {\em Epidemiology}, 23(6):889--891.

\bibitem[Vansteelandt and Joffe, 2014]{Vansteelandt2014}
Vansteelandt, S. and Joffe, M. (2014).
\newblock {Structural Nested Models and G-estimation: The Partially Realized
  Promise}.
\newblock {\em Statistical Science}, 29(4):707--731.

\bibitem[Vermeulen and Vansteelandt, 2015]{Vermeulen2015}
Vermeulen, K. and Vansteelandt, S. (2015).
\newblock {Bias-Reduced Doubly Robust Estimation}.
\newblock {\em Journal of the American Statistical Association},
  110(511):1024--1036.

\bibitem[Wang, 2018]{Wang2018}
Wang, K. (2018).
\newblock {Understanding Power Anomalies in Mediation Analysis}.
\newblock {\em Psychometrika}, 83(2):387--406.

\end{thebibliography}
\bibliographystyle{hapalike} 

\pagebreak
\appendix
\section{Proofs}
\label{appendix_proofs}

\subsection{Proof of Lemma \ref{NIDE_prop_1}}

\label{double_proof_1}
In Step 1 the following expressions for each component of $\E\{U(\beta^*,\gamma^*)\}$ are derived for $\beta^*=(\beta_1,\beta_2,\beta^*_3)$. In Step 2 we consider the behaviour of these expressions in each of the two misspecification cases.
\begin{align}
\E\{U_1(\beta^*,\gamma^*)\} &= \E\left[\left\{ h(Z) - h(Z;\gamma^*_x) \right\}\left\{f(Z)- f(Z;\gamma^*_m)\right\}\right] \label{dp_U1} \\
\E\{U_2(\beta^*,\gamma^*)\} &= \E\left[\left\{ f(Z) - f(Z;\gamma^*_m) \right\} \left\{g(X,Z)- \beta_3^* X  - g(Z;\gamma^*_y) \right\}\right] \label{dp_U2}\\
\E\{U_3(\beta^*,\gamma^*)\} &= \E\left[\left\{ X - h(Z;\gamma^*_x) \right\} \left\{g(X,Z)- \beta_3^* X  - g(Z;\gamma^*_y)  \right\}\right] \label{dp_U3}
\end{align}

Step 1:
For the first component we use the partial linearity to obtain
\begin{align*}
\E\{U_1(\beta^*,\gamma^*)|X,Z\} &= \Big\{X - h(Z;\gamma^*_x)\Big\}\Big\{\E(M-\beta_1 X|X,Z) - f(Z;\gamma^*_m)\Big\}\\
&=  \Big\{X - h(Z;\gamma^*_x)\Big\} \Big\{f(Z) - f(Z;\gamma^*_m)\Big\} \\
\E\{U_1(\beta^*,\gamma^*)|Z\} &= \Big\{h(Z) - h(Z;\gamma^*_x)\Big\}  \Big\{f(Z) - f(Z;\gamma^*_m)\Big\}
\end{align*}
Similarly for the second component,
\begin{align*}
\E\{U_2(\beta^*,\gamma^*)|M,X,Z\} &= \Big\{M-\beta_1 X - f(Z;\gamma^*_m)\Big\} \Big\{\E(Y-\beta_2 M  |M,X,Z)- \beta_3^* X  - g(Z;\gamma^*_y)  \Big\} \\
\E\{U_2(\beta^*,\gamma^*)|X,Z\}   &= \Big\{\E(M-\beta_1 X|X,Z) - f(Z;\gamma^*_m)\Big\} \Big\{g(X,Z)- \beta_3^* X  - g(Z;\gamma^*_y)  \Big\} \\
                                  &= \Big\{f(Z)   - f(Z;\gamma^*_m)\Big\} \Big\{g(X,Z)- \beta_3^* X  - g(Z;\gamma^*_y)  \Big\}
\end{align*}
Finally for the third component,
\begin{align*}
\E\{U_3(\beta^*,\gamma^*)|M,X,Z\} &= \Big\{X - h(Z;\gamma^*_x)\Big\}      \Big\{\E(Y-\beta_2 M  |M,X,Z)- \beta_3^* X  - g(Z;\gamma^*_y) \Big\} \\
\E\{U_3(\beta^*,\gamma^*)|X,Z\}   &= \Big\{X - h(Z;\gamma^*_x)\Big\}      \Big\{g(X,Z)  - \beta_3^* X  - g(Z;\gamma^*_y) \Big\}
\end{align*}

Step 2:
We shall consider the cases (i) and (ii) separately.
In case (i) the conditions for assumption A2 are met, hence $f(Z) = f(Z;\gamma^*_m)$ and so \eqref{dp_U1} and \eqref{dp_U2} are exactly zero. The proof for case (i) is completed by letting $\beta_3^*$ be the value which solves \eqref{dp_U3} equal to zero.

For case (ii) the conditions of A1 are met, hence $h(Z) = h(Z;\gamma^*_x)$ so \eqref{dp_U1} is exactly zero. Also there exists $\beta_3$ such that $g(X,Z)=\beta_3 X + g(Z)$ and for $\beta_3^*=\beta_3$ then the conditions in A3 are met so $g(Z)=g(Z;\gamma^*_y)$ and hence \eqref{dp_U2} and \eqref{dp_U3} are exactly zero, which completes the proof for case (ii).

\subsection{Proof of Lemma \ref{NDE_prop_1}}

\label{double_proof_2}
In Step 1 the following expressions for each component of $\E\{U(\beta^*,\gamma^*)\}$ are derived for $\beta^*=(\beta^*_1,\beta_2,\beta_3)$. In Step 2 we consider the behaviour of these expressions in each of the two misspecification cases.
\begin{align}
\E\{U_1(\beta^*,\gamma^*)\} &= \E\left[\Big\{X - h(Z;\gamma^*_x)\Big\}\Big\{f(X,Z)-\beta_1^*X - f(Z;\gamma^*_m) \Big\} \right] \label{double1_0}\\
\E\{U_2(\beta^*,\gamma^*)\} &= \E\left[ \Big\{f(X,Z) -\beta_1^*X - f(Z;\gamma^*_m)\Big\} \Big\{g(Z)  - g(Z;\gamma^*_y) \Big\} \right] \label{double1_1} \\
\E\{U_3(\beta^*,\gamma^*)\} &= \E\left[ \Big\{h(Z) - h(Z;\gamma^*_x)\Big\}              \Big\{g(Z)  - g(Z;\gamma^*_y) \Big\} \right] \label{double1_2}
\end{align}

Step 1. For the first component,
\begin{align*}
\E\{U_1(\beta^*,\gamma^*)|X,Z\} &= \Big\{X - h(Z;\gamma^*_x)\Big\}      \Big\{f(X,Z)-\beta_1^*X - f(Z;\gamma^*_m) \Big\}
\end{align*}
For the second component we use the partial linearity to obtain
\begin{align*}
\E\{U_2(\beta^*,\gamma^*)|M,X,Z\} &= \Big\{M-\beta_1^*X - f(Z;\gamma^*_m)\Big\} \Big\{\E(Y-\beta_2 M - \beta_3 X |M,X,Z) - g(Z;\gamma^*_y)\Big\} \\
                                  &= \Big\{M-\beta_1^*X - f(Z;\gamma^*_m)\Big\} \Big\{g(Z)     - g(Z;\gamma^*_y)\Big\} \\
\E\{U_2(\beta^*,\gamma^*)|X,Z\}   &= \Big\{f(X,Z) -\beta_1^*X - f(Z;\gamma^*_m)\Big\} \Big\{g(Z) - g(Z;\gamma^*_y)\Big\}
\end{align*}
Similarly for the third component,
\begin{align*}
\E\{U_3(\beta^*,\gamma^*)|M,X,Z\} &= \Big\{X - h(Z;\gamma^*_x)\Big\}       \Big\{\E(Y-\beta_2 M - \beta_3 X |M,X,Z)  - g(Z;\gamma^*_y) \Big\} \\
                                  &= \Big\{X - h(Z;\gamma^*_x)\Big\}       \Big\{g(Z)  - g(Z;\gamma^*_y) \Big\}\\
\E\{U_3(\beta^*,\gamma^*)|Z\}     &= \Big\{h(Z) - h(Z;\gamma^*_x)\Big\} \Big\{g(Z)  - g(Z;\gamma^*_y) \Big\}
\end{align*}

Step 2.
We shall consider the cases (i) and (ii) separately. In case (i) the conditions for assumption A3 are met, hence $g(Z) = g(Z;\gamma^*_y)$ so \eqref{double1_1} and \eqref{double1_2} are exactly zero. Letting $\beta_1^*$ be the value which solves \eqref{double1_0} equal to zero completes the proof for case (i).

For case (ii) the conditions of A1 are met, so $h(Z) = h(Z;\gamma^*_x)$ and so \eqref{double1_2} is zero. Also there exists $\beta_1$ such that $f(X,Z)=\beta_1 X + f(Z)$ and for $\beta_1^*=\beta_1$ then the conditions in A2 are met so $f(Z)=f(Z;\gamma^*_y)$ and hence \eqref{double1_0} and \eqref{double1_1} are exactly zero, which completes the proof for case (ii).

\subsection{Proof of Theorem \ref{double}}

\label{AL_beta}
Here we provide a sketch of the proof. Consider the Taylor Expansion
\begin{align*}
\E_n\{U(\hat{\beta},\hat{\gamma})\} = \E_n\{U(\beta^*,\gamma^*)\} +& \E_n\left\{\frac{\partial U (\beta^*,\gamma^*)}{\partial \beta}\right\} (\hat{\beta} - \beta^*) \\+& \E_n\left\{\frac{\partial U (\beta^*,\gamma^*)}{\partial \gamma}\right\} (\hat{\gamma} - \gamma^*) + o_p\left( n^{-1/2}\right)
\end{align*}
Since $\E_n\{U(\hat{\beta},\hat{\gamma})\} = 0$ then
\begin{align*}
\hat{\beta} - \beta^* &= \E_n\left\{-\frac{\partial U (\beta^*,\gamma^*)}{\partial \beta}\right\}^{-1} \left[ \E_n\{U(\beta^*,\gamma^*)\}  + \E_n\left\{\frac{\partial U (\beta^*,\gamma^*)}{\partial \gamma}\right\} (\hat{\gamma} - \gamma^*) \right] + o_p\left( n^{-1/2}\right) \\
\end{align*}
Using the estimator in \eqref{nuis_estimator} and rearranging gives
\begin{align*}
\hat{\beta} - \beta^*  &= \E_n\left( \E_n\left\{-\frac{\partial U (\beta^*,\gamma^*)}{\partial \beta}\right\}^{-1} \left[ U(\beta^*,\gamma^*)  + \E_n\left\{\frac{\partial U (\beta^*,\gamma^*)}{\partial \gamma}\right\} \phi(\beta^*,\gamma^*) \right] \right) + o_p\left( n^{-1/2}\right)
\end{align*}
Applying the weak law of large numbers to the partial derivative terms gives the form of the influence function $\varphi(.)$ in \eqref{beta_IF}. We must further show that $\E\{\varphi(\beta^*,\gamma^*)\} = 0$
\begin{align*}
\E\{\varphi(\beta^*,\gamma^*)\} &= \E\left\{-\frac{\partial U (\beta^*,\gamma^*)}{\partial \beta}\right\}^{-1} \left[ \E\{U(\beta^*,\gamma^*)\}  + \E\left\{\frac{\partial U (\beta^*,\gamma^*)}{\partial \gamma}\right\} \E\{\phi(\beta^*,\gamma^*)\} \right]
\end{align*}
Since $\phi(.)$ is an influence function, $\E\{\phi(\beta^*,\gamma^*)\} = 0$. Therefore provided $\E\{U(\beta^*,\gamma^*)\} = 0$ then $\E\{\varphi(\beta^*,\gamma^*)\} = 0$ as required.

\subsection{Derivation of Equation \eqref{eta_infl}}

\label{eta_proof}
By Theorem \ref{double},
\begin{align*}
\hat{\beta}_1 &= \beta_1 + \E_n\{\varphi_1(\beta^*,\gamma^*)\}  +o_p\left( n^{-1/2}\right) \\
\hat{\beta}_2 &= \beta_2 + \E_n\{\varphi_2(\beta^*,\gamma^*)\}  +o_p\left( n^{-1/2}\right)
\end{align*}
Therefore, letting $A = \E_n\{\varphi_1(\beta^*,\gamma^*)\}$ and $B = \E_n\{\varphi_2(\beta^*,\gamma^*)\}$,
\begin{align*}
\hat{\beta}_1\hat{\beta}_2 - \beta_1\beta_2 &= \E_n\{\omega(\beta^*,\gamma^*)\} + AB  + o_p\left( n^{-1/2}\right)
\end{align*}
and the desired result follows provided that $AB = o_p\left( n^{-1/2}\right)$. Using Markov's inequality,
\begin{align*}
\pr(|n^{1/2}AB|\geq \epsilon) = \pr(n(AB)^2 \geq \epsilon^2) \leq \frac{n\E\{(AB)^2\}}{\epsilon^2}
\end{align*}
Examining the expectation term, we find a sum over four indices
\begin{align*}
\E\{(AB)^2\} = n^{-4}\sum_{i=1}^n \sum_{j=1}^n \sum_{k=1}^n \sum_{l=1}^n \E\{\varphi_1^{(i)}(\beta^*,\gamma^*)\varphi_1^{(j)}(\beta^*,\gamma^*)\varphi_2^{(k)}(\beta^*,\gamma^*)\varphi_2^{(l)}(\beta^*,\gamma^*)\}
\end{align*}
where the superscript $(i)$ denotes that the influence function is evaluated on the $i$th observation. Since the observations are iid and the influence function has mean zero, the terms of this quadruple sum can only be non-zero when their indices are paired, i.e when ($i=j$ and $k=l$) or ($i=k$ and $j=l$) or ($i=l$ and $j=k$). The number of non-zero terms in the sum is therefore of order $n^2$, and hence
\begin{align*}
\pr(|n^{1/2}AB|\geq \epsilon) \leq \mathcal{O} \left(n^{-1}\right)
\end{align*}
where $\mathcal{O}$ denotes conventional big-O notation, i.e. for sufficiently large $n$ there exists some constant $k$ such that $|\mathcal{O} \left(n^{-1}\right)| \leq k n^{-1}$

\subsection{Proof of Theorem \ref{dufour_theorem}}

Here we adapt the proof from Section 5.1 of \cite{Dufour2017} to allow for orthogonal nuisance parameter estimation. We prove the results for the CUE estimator, however they are equally applicable to the two-step estimator. Our extension to the original results relies on three orthogonality-like derivative results for the test statistic of interest. These are derived assuming that the nuisance parameter estimator is orthogonal to the moment conditions in the sense that \eqref{Neyman_orthog} holds. This may either be because all models are correctly specified or because a bias-reduced strategy is used to estimate nuisance parameters, as described below. We begin by defining the CUE objective function, which we denote by $M_n$ as in the original notation of \cite{Dufour2017},
\begin{align*}
M_n(\beta,\gamma) &= D_n^\top (\beta,\gamma)I^{-1}_n(\beta,\gamma)D_n (\beta,\gamma)
\end{align*}
where, for a target parameter moment function $U(\beta,\gamma)$,
\begin{align*}
D_n (\beta,\gamma) &= \E_n[U(\beta,\gamma)] \\
C_n (\beta,\gamma) &= \E_n\left[\frac{\partial U(\beta,\gamma)}{\partial \gamma}\right] = \frac{\partial D_n (\beta,\gamma) }{\partial \gamma}\\
I_n(\beta,\gamma) &= \E_n[U(\beta,\gamma)U(\beta,\gamma)^\top]
\end{align*}
Theorem \ref{dufour_theorem} in the text considers the exactly specified setting, i.e. $\Dim(\beta)=\Dim(U(\beta,\gamma))$. For the current proof, however, we consider the over-specified setting, i.e. $\Dim(\beta) \leq \Dim(U(\beta,\gamma))$.
Also define the probability limits, $\beta^*,\gamma^*$ as the (assumed to be) unique values such that
\begin{align*}
\frac{\partial M_n(\beta^*,\gamma^*) }{ \partial \beta}  &\overset{p}{\to} 0 \\
C_n (\beta^*,\gamma^*) &\overset{p}{\to} 0
\end{align*}
The first of these is equivalent to $D_n (\beta^*,\gamma^*) \overset{p}{\to} 0$ in the exactly specified setting. By the central limit theorem,
\begin{align*}
\sqrt{n} D_n(\beta^*,\gamma^*) &\overset{d}{\to}  \n{0}{I_0} \\
I_n(\beta^*,\gamma^*) &\overset{p}{\to} I_0 = \E[U(\beta^*,\gamma^*)U(\beta^*,\gamma^*)^\top]
\end{align*}
The unconstrained estimated values $\hat{\beta}, \hat{\gamma}$ are those which solve
\begin{align}
\frac{\partial M_n(\hat{\beta},\hat{\gamma}) }{ \partial \beta}  &= 0 \label{est1}\\
C_n (\hat{\beta},\hat{\gamma}) &= 0 \nonumber
\end{align}
Again, the first of these is equivalent to $D_n (\hat{\beta},\hat{\gamma}) = 0$ in the exactly specified setting. The constrained estimated values $\hat{\beta}_\psi, \hat{\gamma}_\psi$ are those which solve
\begin{align}
\frac{\partial M_n(\hat{\beta}_\psi,\hat{\gamma}_\psi) }{ \partial \beta} - \frac{\partial \psi(\hat{\beta}_\psi)}{\partial \beta} \lambda &= 0 \label{est2}\\
C_n (\hat{\beta}_\psi,\hat{\gamma}_\psi)  &= 0 \nonumber \\
\psi(\hat{\beta}_\psi) &= 0 \label{est3}
\end{align}
for a constraint function $\psi$ and where $\lambda$ is a Lagrange multiplier. The statement that we intend to prove is that
\begin{align}
n[M_n(\hat{\beta}_\psi,\hat{\gamma}_\psi) - M_n(\hat{\beta},\hat{\gamma})] \overset{d}{\to} \chi^2_r
\end{align}
where $r$ is the rank of $\partial \psi(\beta)/\partial\beta$ in a neighbourhood of $\beta^*$. In the exactly specified setting, $M_n(\hat{\beta},\hat{\gamma}) = 0$.

\subsubsection*{Three necessary derivative results}

In this subsection we show that, since $C_n (\beta^*,\gamma^*) = o_p(1), D_n (\beta^*,\gamma^*) = o_p(1)$, and $\sqrt{n} D_n (\beta^*,\gamma^*) = O_p(1)$ then
\begin{align}
\sqrt{n}\frac{\partial M_n(\beta^*,\gamma^*) }{\partial \gamma} &= o_p(1) \label{deriv_1} \\
\frac{\partial^2 M_n(\beta^*,\gamma^*) }{\partial \gamma \partial \beta} &= o_p(1) \label{deriv_2} \\
\frac{\partial^2 M_n(\beta^*,\gamma^*) }{\partial \gamma \partial \gamma} &= o_p(1) \label{deriv_3}
\end{align}
To do so it is easier to work in an index notation where $D_i$ is the $ith$ component of $D_n(\beta,\gamma)$, and $I^{-1}_{ij}$ is the $i,jth$ component of $I_n^{-1}(\beta,\gamma)$ and all quantities are evaluated at $(\beta,\gamma) = (\beta^*,\gamma^*)$. For example, letting $q=\Dim(U(\beta,\gamma))$, then
\begin{align*}
M_n(\beta^*,\gamma^*) =  \sum_{i=1}^{q}\sum_{j=1}^{q} D_i I^{-1}_{ij} D_j
\end{align*}
For the first derivative term of interest,
\begin{align*}
\sqrt{n}\frac{\partial M_n(\beta^*,\gamma^*) }{\partial \gamma} &= \sum_{i=1}^{q}\sum_{j=1}^{q} \left\{ \left(2\frac{\partial D_i }{\partial \gamma} I^{-1}_{ij} +   D_i \frac{\partial I^{-1}_{ij}}{\partial \gamma}\right) \sqrt{n} D_j \right\}
\end{align*}
and for the second derivative term of interest,
\begin{align*}
\frac{\partial^2 M_n(\beta^*,\gamma^*) }{\partial \gamma \partial \beta} &=  \sum_{i=1}^{q}\sum_{j=1}^{q} \left\{ 2 \frac{\partial D_i}{\partial \beta} I^{-1}_{ij} \frac{\partial D_j}{\partial \gamma} + \left( 2 \frac{\partial^2 D_i}{\partial \gamma \partial \beta} I^{-1}_{ij} + 2 \frac{\partial D_i}{\partial \beta} \frac{\partial I^{-1}_{ij}}{\partial \gamma} +  2 \frac{\partial D_i}{\partial \gamma} \frac{\partial I^{-1}_{ij}}{\partial \beta} + D_i \frac{\partial^2 I^{-1}_{ij}}{\partial \gamma \partial \beta} \right)D_j  \right\}
\end{align*}
For the third,
\begin{align*}
\frac{\partial^2 M_n(\beta^*,\gamma^*) }{\partial \gamma \partial \gamma} &= \sum_{i=1}^{q}\sum_{j=1}^{q} \left\{ 2 \frac{\partial D_i}{\partial \gamma} I^{-1}_{ij} \frac{\partial D_j}{\partial \gamma} + \left( 2 \frac{\partial^2 D_i}{\partial \gamma \partial \gamma} I^{-1}_{ij} + 2 \frac{\partial D_i}{\partial \beta} \frac{\partial I^{-1}_{ij}}{\partial \gamma} +  2 \frac{\partial D_i}{\partial \gamma} \frac{\partial I^{-1}_{ij}}{\partial \gamma} + D_i \frac{\partial^2 I^{-1}_{ij}}{\partial \gamma \partial \gamma} \right) D_j \right\}
\end{align*}
By the orthogonality of the nuisance parameter estimator, $\partial D_j/\partial \gamma = o_p(1)$, and since $D_j = o_p(1)$, and $\sqrt{n} D_j= O_p(1)$ then the results follow.

\subsubsection*{Applying the derivative results}

Consider the test statistic
\begin{align*}
\xi = M_n(\hat{\beta}_\psi,\hat{\gamma}_\psi) - M_n(\hat{\beta},\hat{\gamma})
\end{align*}
Under standard regularity conditions, and the rank condition in \eqref{rank:cond} (see \cite{Dufour2017} for details),  $\hat{\gamma},\hat{\gamma}_\psi,\hat{\beta}$ and $\hat{\beta}_\psi$ are CAN, hence expanding this test statistics to second order gives
\begin{align*}
n\xi &= \sqrt{n}\frac{\partial M_n(\beta^*,\gamma^*)}{\partial \beta^\top} \left\{ \sqrt{n}(\hat{\beta}_\psi - \beta^*) - \sqrt{n}(\hat{\beta} - \beta^*) \right\} \\
&+ \sqrt{n}\frac{\partial M_n(\beta^*,\gamma^*)}{\partial \gamma^\top} \left\{ \sqrt{n}(\hat{\gamma}_\psi - \gamma^*) - \sqrt{n}(\hat{\gamma} - \gamma^*) \right\} \\
&+ \frac{1}{2} \left\{ \sqrt{n}(\hat{\beta}_\psi - \beta^*)^\top \frac{\partial^2 M_n(\beta^*,\gamma^*) }{ \partial \beta \partial \beta^\top}  \sqrt{n}(\hat{\beta}_\psi - \beta^*)^\top - \sqrt{n}(\hat{\beta} - \beta^*)^\top \frac{\partial^2 M_n(\beta^*,\gamma^*) }{ \partial \beta \partial \beta^\top}  \sqrt{n}(\hat{\beta} - \beta^*)^\top \right\} \\
&+ \frac{1}{2} \left\{ \sqrt{n}(\hat{\gamma}_\psi - \gamma^*)^\top \frac{\partial^2 M_n(\beta^*,\gamma^*) }{ \partial \gamma \partial \gamma^\top}  \sqrt{n}(\hat{\gamma}_\psi - \gamma^*)^\top - \sqrt{n}(\hat{\gamma} - \gamma^*)^\top \frac{\partial^2 M_n(\beta^*,\gamma^*) }{ \partial \gamma \partial \gamma^\top}  \sqrt{n}(\hat{\gamma} - \gamma^*)^\top \right\}\\
&+ \left\{ \sqrt{n}(\hat{\gamma}_\psi - \gamma^*)^\top \frac{\partial^2 M_n(\beta^*,\gamma^*) }{ \partial \gamma \partial \beta^\top}  \sqrt{n}(\hat{\beta}_\psi - \beta^*)^\top - \sqrt{n}(\hat{\gamma} - \gamma^*)^\top \frac{\partial^2 M_n(\beta^*,\gamma^*) }{ \partial \gamma \partial \beta^\top}  \sqrt{n}(\hat{\beta} - \beta^*)^\top \right\} \\
&+ o_p(1)
\end{align*}
Using the derivative results \eqref{deriv_1} to \eqref{deriv_3} our expansion reduces to
\begin{align*}
n\xi &= \sqrt{n}\frac{\partial M_n(\beta^*,\gamma^*)}{\partial \beta^\top} \left\{ \sqrt{n}(\hat{\beta}_\psi - \beta^*) - \sqrt{n}(\hat{\beta} - \beta^*) \right\} \\
&+ \frac{1}{2} \left\{ \sqrt{n}(\hat{\beta}_\psi - \beta^*)^\top \frac{\partial^2 M_n(\beta^*,\gamma^*) }{ \partial \beta \partial \beta^\top}  \sqrt{n}(\hat{\beta}_\psi - \beta^*)^\top - \sqrt{n}(\hat{\beta} - \beta^*)^\top \frac{\partial^2 M_n(\beta^*,\gamma^*) }{ \partial \beta \partial \beta^\top}  \sqrt{n}(\hat{\beta} - \beta^*)^\top \right\} \\
&+ o_p(1)
\end{align*}

Next we consider the first order Taylor expansions of the estimating equations \eqref{est1} to \eqref{est3}, taken about the probability limit values. Again, since $\hat{\gamma},\hat{\gamma}_\psi,\hat{\beta}$ and $\hat{\beta}_\psi$ are CAN,
\begin{align*}
0 &= \frac{\partial M_n(\beta^*,\gamma^*) }{ \partial \beta} + \frac{\partial^2 M_n(\beta^*,\gamma^*) }{ \partial \beta \partial \beta} (\hat{\beta} - \beta^*) + \frac{\partial^2 M_n(\beta^*,\gamma^*) }{ \partial \gamma \partial \beta} (\hat{\gamma} - \gamma^*) + o_p(n^{-1/2})\\
0 &= \frac{\partial M_n(\beta^*,\gamma^*) }{ \partial \beta} + \frac{\partial^2 M_n(\beta^*,\gamma^*) }{ \partial \beta \partial \beta} (\hat{\beta}_\psi - \beta^*) + \frac{\partial^2 M_n(\beta^*,\gamma^*) }{ \partial \gamma \partial \beta} (\hat{\gamma}_\psi - \gamma^*) - \frac{\partial \psi(\hat{\beta}_\psi)}{\partial \beta} \lambda + o_p(n^{-1/2})\\
0 &= \psi(\beta^*) +  \frac{\partial \psi(\beta^*)}{\partial \beta} (\hat{\beta}_\psi - \beta^*) + o_p(n^{-1/2})
\end{align*}
Under the null, $\psi(\beta^*) = 0$, application of \eqref{deriv_2} gives
\begin{align*}
0 &= X_n + V_0 (\hat{\beta} - \beta^*)  + o_p(n^{-1/2})\\
0 &= X_n + V_0 (\hat{\beta}_\psi - \beta^*) - P_0 \lambda + o_p(n^{-1/2}) \\
0 &= P_0 (\hat{\beta}_\psi - \beta^*) + o_p(n^{-1/2})
\end{align*}
where,
\begin{align*}
\frac{\partial \psi(\beta^*)}{\partial \beta} &= P_0 \\
\frac{\partial M_n(\beta^*,\gamma^*) }{ \partial \beta} &= X_n\\
\frac{\partial^2 M_n(\beta^*,\gamma^*) }{ \partial \beta \partial \beta} &\overset{p}{\to} V_0 
\end{align*}
It follows immediately from the original proof in \cite{Dufour2017} that
\begin{align*}
\xi = \frac{1}{2} X_n^\top V_0^{-1}P_0^\top(P_0V_0^{-1}P_0^\top)^{-1}P_0V_0^{-1}X_n + o_p(n^{-1})
\end{align*}
The final result follows when $\sqrt{n}X_n \overset{d}{\to} \n{0}{2V_0}$. This can be shown using the same derivative methods as used to show \eqref{deriv_1} to \eqref{deriv_3}.

\subsection{Proof of Equations \eqref{LR_result}  and \eqref{score_result}}

\label{proof_valid}
We will prove \eqref{LR_result} with the result for \eqref{score_result} proceeding in a similar fashion.

Consider Theorem \ref{dufour_theorem} under the null hypothesis $\psi^{(0)}(\beta^*) = \beta_1\beta_2 = 0$. With the null parameter space given by $B_0 = \{\beta|\psi^{(0)}(\beta) = 0\}$, with
\begin{align*}
\Rank \left(\frac{\partial \psi^{(0)}(\beta)}{\partial \beta} \right)  = \begin{cases}
0 & \text{for } \beta_1 = \beta_2 = 0 \\
1 & \text{otherwise }
\end{cases}
\end{align*}
One may  decompose the supremum in \eqref{LR_result} as
\begin{align}
\sup_{\beta_* \in B_0} \pr_{\beta^*} \left( S_0 > x  \right) &= \max\left\{\sup_{\beta^* \in B_0\setminus A} \pr_{\beta^*} \left( S_0 > x  \right) , \sup_{\beta^* \in A} \pr_{\beta^*} \left( S_0 > x  \right)\right\} \label{Eq.proof_valid}
\end{align}
where $A = \{\beta|\beta_1=\beta_2=0\}$.
For the first term in the max bracket above, the rank condition of Theorem \ref{dufour_theorem} holds, so for all $\beta^* \in B_0\setminus A$
\begin{align*}
\pr_{\beta^*} \left( S_0 > x \right) \to 1 - F_{\chi^2_1} (x)
\end{align*}
Considering the second term, one may decompose the test statistic as
\begin{align*}
\pr_{\beta^*} \left( S_0 > x  \right)= \pr_{\beta^*} \left( S_1 > x, S_2 > x  \right)\leq \pr_{\beta^*} \left( S_1 > x\right)
\end{align*}
where (for $j=1,2$) $S_j = \min_{\beta \in C_j} S(\beta)$ and $C_j = \{\beta|\beta_j=0\}$. By Theorem \ref{dufour_theorem}, for all $\beta^*$ in $A$,
\begin{align*}
\pr_{\beta^*} \left( S_1 > x\right) \to 1 - F_{\chi^2_1} (x)
\end{align*}
Hence $\pr_{\beta^*} \left( S_0 > x  \right)$ is asymptotically bounded from above by  $1 - F_{\chi^2_1} (x)$ for all $\beta^*$ in $B_0$, so \eqref{LR_result} holds.

\subsection{G-estimation when outcome model has exposure-mediator interaction}

In the following we reason about the NIDE obtained by G-estimation using moment conditions \eqref{G1} -- \eqref{G3}, when one has erroneously excluded an interaction term from the outcome model, but the mediator model, $\E(M|X,Z)$ is correctly specified and partially linear, i.e. in truth, \eqref{m.mod} and \eqref{y.mod.int} both hold. We define the probability limit as $\beta^* = (\beta_1^*,\beta_2^*,\beta_3^*)$ which solves $\E\{U(\beta^*,\gamma^*)\} = 0$ and use assumption A2 as before. Let,
\begin{align*}
\delta_j &= \beta_j - \beta_j^* \\
\epsilon_x &= X - \E(X|Z) \\
\epsilon_m &= M - \E(M|X,Z) \\
\beta_3 &= \frac{\E(\epsilon_x g(X,Z))}{\E(\epsilon_x X)} \\
\Delta_f &= f(Z) - f(Z;\gamma_m^*) \\
\Delta_g &= g(X,Z) - \beta_3 X - g(Z;\gamma_m^*) \\
\Delta_h &= h(Z) - h(Z;\gamma_m^*)
\end{align*}
for $j=1,2,3$. Here $\beta_3$ is the least squares coefficient of a regression of $g(X,Z)$ on $X$. It follows that the expected moment conditions can be written
\begin{align*}
\E[U_1(\beta^*,\gamma^*)] &= \E\{ (\epsilon_x + \Delta_h) (\delta_1 X + \Delta_f) \}\\
\E[U_2(\beta^*,\gamma^*)] &= \E\{ (\epsilon_m + \delta_1 X + \Delta_f) (\delta_2 M + \theta MX + \delta_3 X + \Delta_g ) \} \\
\E[U_3(\beta^*,\gamma^*)] &= \E\{ (\epsilon_x + \Delta_h) (\delta_2 M + \theta MX + \delta_3 X + \Delta_g ) \}
\end{align*}
For the first equation, since $\E(\epsilon_x \Delta_f) = 0$, then
\begin{align*}
\E[U_1(\beta^*,\gamma^*)] &= \delta_1 \E[ (\epsilon_x + \Delta_h)X ] + \E (\Delta_h \Delta_f)
\end{align*}
We assume that $f(z)$ is modelled correctly and when $\delta_1=0$ then assumption A2 is satisfied and $\Delta_f = 0$. Using this fact, the second equation becomes
\begin{align*}
\E[U_2(\beta^*,\gamma^*)] &= \delta_2 \E( \epsilon_m M ) + \theta \E( \epsilon_m MX )
\end{align*}
where we have used the fact that $\E(\epsilon_m X) = \E(\epsilon_m \Delta_g) = 0$. Hence,
\begin{align*}
\delta_2 &= -\theta \frac{\E(\epsilon_m MX)}{\E(\epsilon_m M)}
\end{align*}
which, since $\delta_1=0$, gives the result
\begin{align*}
\beta_1^*\beta_2^* = \beta_1 \left(\beta_2 + \theta \bar{x} \right)
\end{align*}
where
\begin{align}
\bar{x} = \frac{\E(\epsilon_m MX)}{\E(\epsilon_m M)} =  \frac{\E[X\mathrm{var}(M|X,Z)]}{\E[\mathrm{var}(M|X,Z)]}
\end{align}
can be thought of as a weighted average of  $X$, or as a population least squares regression coefficient from regressing $MX$ on $M$.

\section{Additional Simulation Plots}
\label{add_plots}
\pagebreak
\begin{figure}[tbp]
\centering
\includegraphics[scale=0.9]{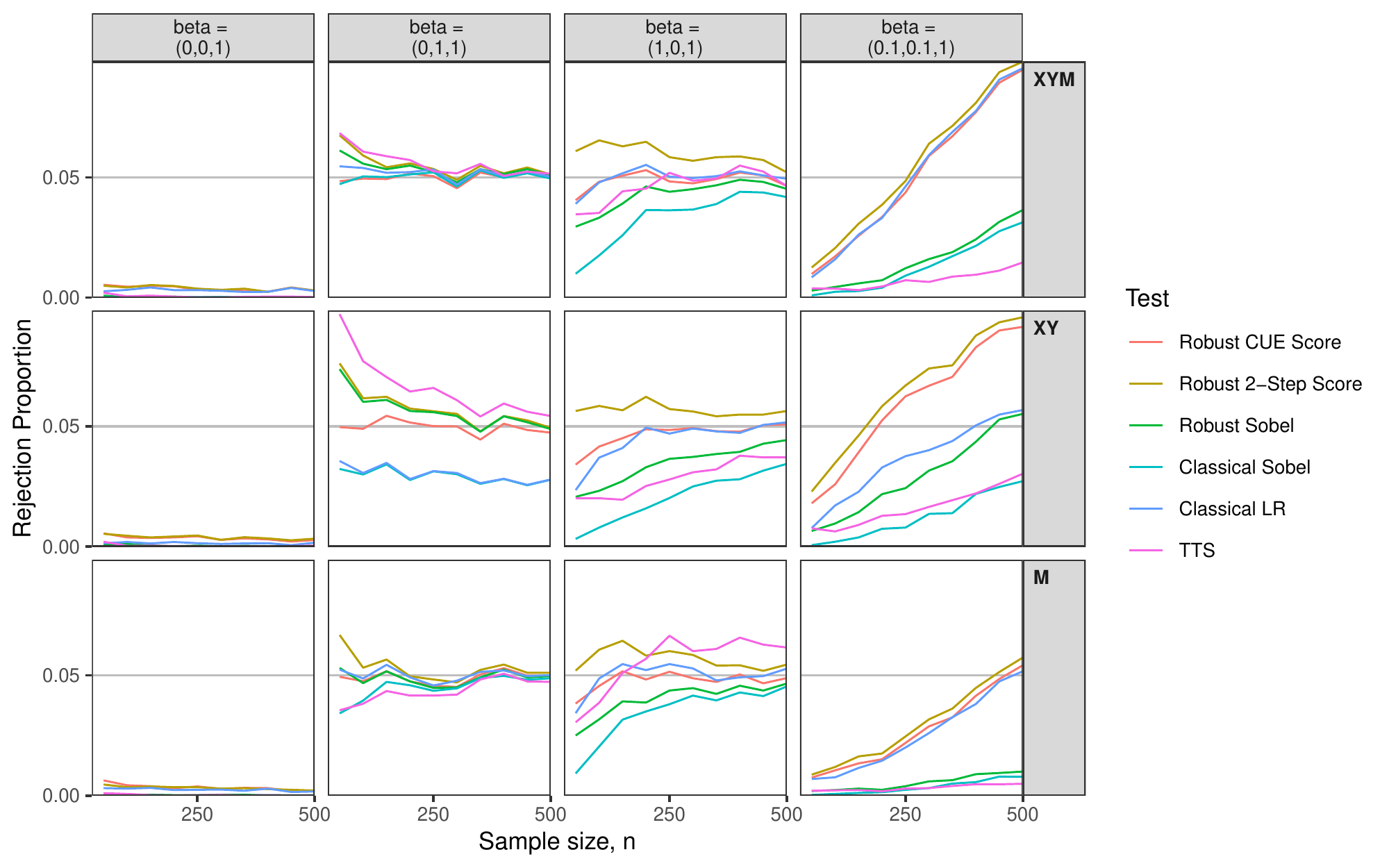}
\caption{Simulated results from data generating process B of the proportion of the $10^4$ datasets for which the no-mediation hypothesis ($H_0$) is rejected at the 5\% level testing using the CUE score, Two-step score, Robust Sobel, Classical Sobel, Classical LR, and TTS methods. Each column represents a different true $\beta$ parameter, whilst each row gives the models which are correctly specified (those for which the misspecification indicator is equal to zero)}
\label{sim_plot_1_t}
\end{figure}

\begin{figure}[tbp]
\centering
\includegraphics[scale=0.9]{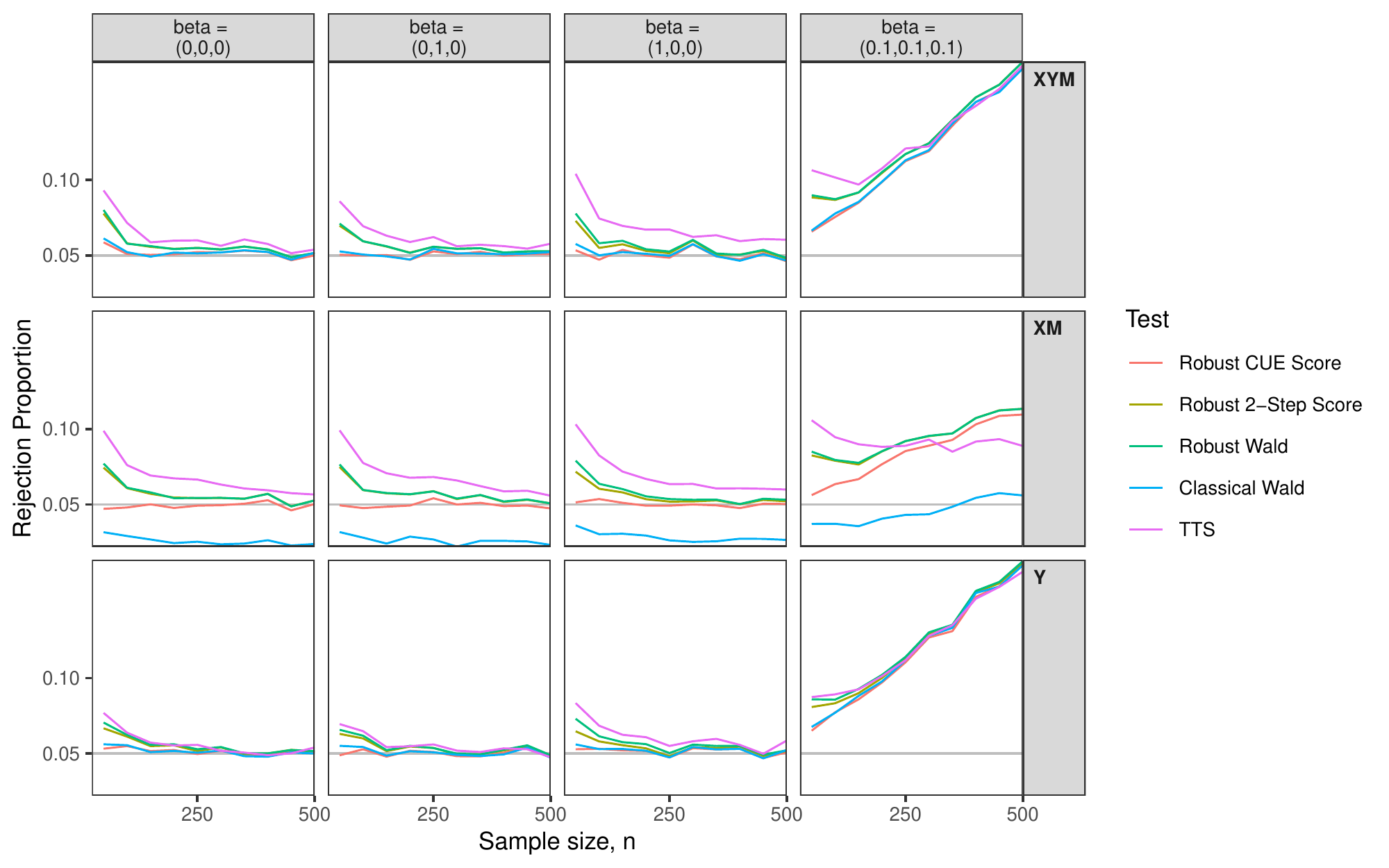}
\caption{Simulated results from data generating process B of the proportion of the $10^4$ datasets for which the no-direct effect hypothesis ($H_1$) is rejected at the 5\% level testing using the CUE score, Two-step score, Robust Wald, Classical Wald, and TTS methods. Each column represents a different true $\beta$ parameter, whilst each row gives the models which are correctly specified (those for which the misspecification indicator is equal to zero)}
\label{sim_plot_2_t}
\end{figure}

\begin{figure}[tbp]
\centering
\includegraphics[scale=0.9]{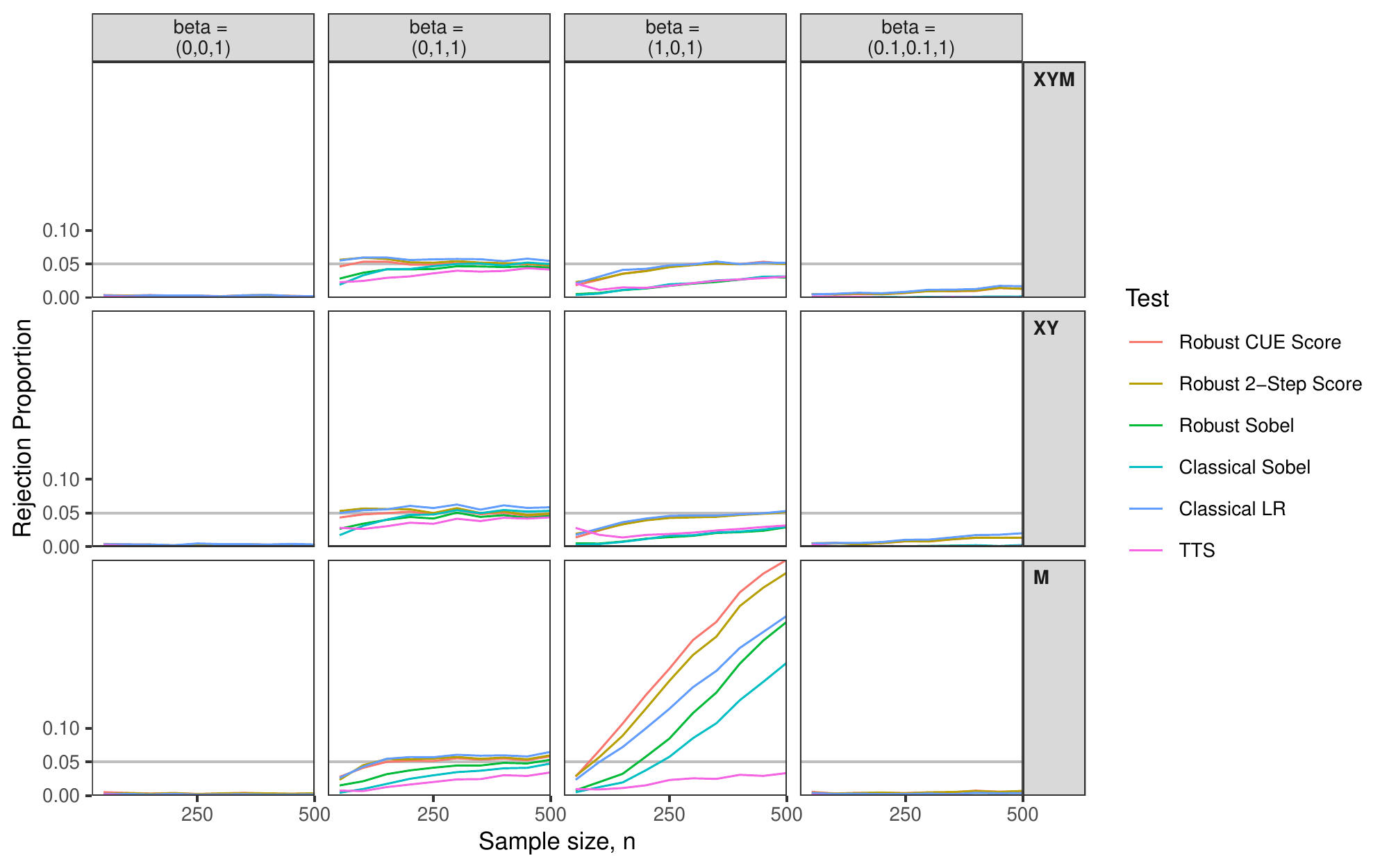}
\caption{Simulated results from data generating process C of the proportion of the $10^4$ datasets for which the no-mediation hypothesis ($H_0$) is rejected at the 5\% level testing using the CUE score, Two-step score, Robust Sobel, Classical Sobel, Classical LR, and TTS methods. Each column represents a different true $\beta$ parameter, whilst each row gives the models which are correctly specified (those for which the misspecification indicator is equal to zero)}
\label{sim_plot_1_log}
\end{figure}

\begin{figure}[tbp]
\centering
\includegraphics[scale=0.9]{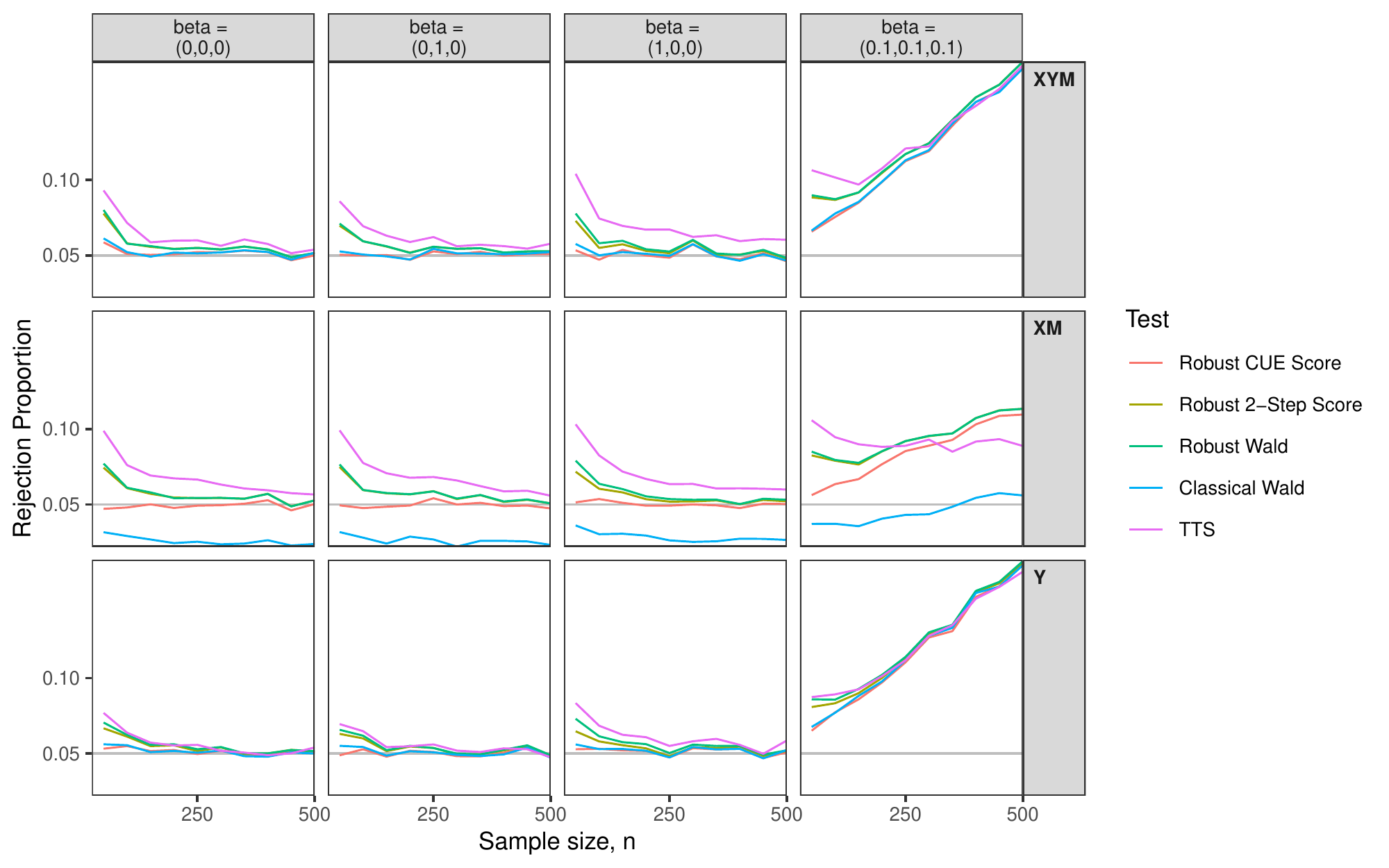}
\caption{Simulated results from data generating process C of the proportion of the $10^4$ datasets for which the no-direct effect hypothesis ($H_1$) is rejected at the 5\% level testing using the CUE score, Two-step score, Robust Wald, Classical Wald, and TTS methods. Each column represents a different true $\beta$ parameter, whilst each row gives the models which are correctly specified (those for which the misspecification indicator is equal to zero)}
\label{sim_plot_2_log}
\end{figure}

\section{Simulation Data Tables}
\label{add_tables}

The Tables \ref{nde_dgpA} to \ref{nide_dgpC} present the results of the bias simulation where $10^3$ dataset replicates were used to estimate the bias and variance of the G-estimator and the TTS based estimators. The variance of the G-estimators were calculated in 3 different ways, using the empirical variance over the $10^3$ data set replicates, using theoretical results presented in this paper and using Bootstrap with $10^3$ bootstrap iterations. For the TTS methods only the empirical variance was calculated. All variances estimates have been scaled by the sample size $N$. Theoretical and Bootstrap estimates are given as the mean over the $10^3$ dataset replicates. The standard error of each quantity is given in brackets.

\pagestyle{plain}

\begin{landscape}
\centering
\begin{longtable}{|c|c|c|c|c|c|c|c|c|c|}
\caption{NDE estimand for Data Generating Process A}
\label{nde_dgpA}
\endfirsthead
\endhead
\hline
Beta & N & Spec & Bias\_G & Bias\_TTS & Var\_G & Var\_TTS & Theory\_Var\_G & Boot\_Var\_G \\ 
  \hline
(0,0,0) & 100 & XYM & 0.00205(0.00718) & -0.00188(0.0077) & 5.15(0.23) & 5.92(0.265) & 4.76(0.0327) & 4.97(0.035) \\ 
  (0,0,0) & 100 & XY & -0.000128(0.00708) & -0.00336(0.00737) & 5.01(0.224) & 5.44(0.243) & 4.75(0.0327) & 4.94(0.0347) \\ 
  (0,0,0) & 100 & XM & 0.0171(0.00974) & 0.0147(0.016) & 9.49(0.425) & 25.6(1.15) & 9.37(0.115) & 9.64(0.111) \\ 
  (0,0,0) & 100 & M & 0.843(0.0104) & 0.932(0.0112) & 10.9(0.488) & 12.5(0.558) & 10.2(0.108) & 10.6(0.111) \\ 
  (0,0,0) & 100 & Y & -0.00524(0.00728) & -0.0053(0.00738) & 5.3(0.237) & 5.45(0.244) & 4.72(0.0299) & 4.89(0.0319) \\ 
  (0,0,0) & 500 & XYM & 0.00409(0.00311) & 0.00318(0.00327) & 4.83(0.216) & 5.35(0.239) & 4.84(0.0142) & 4.88(0.0156) \\ 
  (0,0,0) & 500 & XY & 0.00205(0.00308) & 0.00241(0.0032) & 4.73(0.212) & 5.12(0.229) & 4.82(0.0133) & 4.84(0.0149) \\ 
  (0,0,0) & 500 & XM & 0.00245(0.00443) & 0.00352(0.00791) & 9.8(0.438) & 31.3(1.4) & 9.69(0.0554) & 9.73(0.057) \\ 
  (0,0,0) & 500 & M & 0.861(0.00444) & 0.941(0.00456) & 9.88(0.442) & 10.4(0.464) & 10.4(0.0487) & 10.5(0.0499) \\ 
  (0,0,0) & 500 & Y & 0.0026(0.00316) & 0.00141(0.00321) & 4.99(0.223) & 5.14(0.23) & 4.79(0.0125) & 4.82(0.0143) \\ 
  (0,0,0) & 1000 & XYM & -0.00136(0.00221) & -0.00103(0.00227) & 4.87(0.218) & 5.16(0.231) & 4.85(0.00996) & 4.86(0.0123) \\ 
  (0,0,0) & 1000 & XY & 0.000526(0.00218) & 0.000912(0.00226) & 4.76(0.213) & 5.09(0.228) & 4.83(0.00941) & 4.83(0.0113) \\ 
  (0,0,0) & 1000 & XM & -0.000147(0.00309) & 0.00113(0.0051) & 9.57(0.428) & 26.1(1.17) & 9.58(0.038) & 9.59(0.0395) \\ 
  (0,0,0) & 1000 & M & 0.869(0.00331) & 0.95(0.0034) & 11(0.491) & 11.6(0.518) & 10.6(0.0383) & 10.7(0.0406) \\ 
  (0,0,0) & 1000 & Y & 0.00184(0.00216) & 0.00153(0.00221) & 4.66(0.208) & 4.88(0.218) & 4.8(0.00878) & 4.82(0.0113) \\ 
  (1,1,1) & 100 & XYM & 0.0107(0.00776) & 0.00396(0.0103) & 6.03(0.27) & 10.5(0.471) & 5.82(0.0437) & 6.11(0.0459) \\ 
  (1,1,1) & 100 & XY & -0.00412(0.00745) & -0.00188(0.00778) & 5.55(0.249) & 6.05(0.271) & 5.13(0.036) & 5.39(0.0388) \\ 
  (1,1,1) & 100 & XM & -0.0089(0.0112) & -0.0113(0.0193) & 12.4(0.557) & 37.4(1.67) & 12.2(0.152) & 12.7(0.152) \\ 
  (1,1,1) & 100 & M & 0.838(0.0115) & 0.914(0.0159) & 13.3(0.596) & 25.2(1.13) & 12.8(0.145) & 13.4(0.149) \\ 
  (1,1,1) & 100 & Y & -0.0135(0.0073) & -0.00428(0.00813) & 5.33(0.238) & 6.61(0.296) & 5.7(0.0387) & 5.95(0.0407) \\ 
  (1,1,1) & 500 & XYM & 0.00438(0.00351) & 0.00847(0.00445) & 6.14(0.275) & 9.92(0.444) & 5.85(0.0192) & 5.9(0.0211) \\ 
  (1,1,1) & 500 & XY & 0.00309(0.00316) & 0.00385(0.0033) & 4.99(0.223) & 5.44(0.243) & 5.16(0.0151) & 5.21(0.0173) \\ 
  (1,1,1) & 500 & XM & -0.00834(0.00511) & -0.00323(0.0115) & 13(0.583) & 66.5(2.98) & 12.6(0.0688) & 12.7(0.0705) \\ 
  (1,1,1) & 500 & M & 0.872(0.00511) & 0.954(0.00691) & 13(0.584) & 23.9(1.07) & 13.3(0.0709) & 13.4(0.0732) \\ 
  (1,1,1) & 500 & Y & 0.00428(0.00336) & 0.0034(0.00358) & 5.64(0.252) & 6.42(0.287) & 5.7(0.0161) & 5.75(0.0182) \\ 
  (1,1,1) & 1000 & XYM & 4.3e-05(0.00237) & -0.000762(0.00309) & 5.6(0.251) & 9.53(0.426) & 5.81(0.0138) & 5.85(0.0163) \\ 
  (1,1,1) & 1000 & XY & -8.16e-05(0.00223) & 0.000244(0.00232) & 4.98(0.223) & 5.37(0.24) & 5.16(0.0109) & 5.2(0.0131) \\ 
  (1,1,1) & 1000 & XM & 0.00194(0.00361) & 0.0103(0.00841) & 13(0.583) & 70.7(3.16) & 12.7(0.0531) & 12.7(0.0564) \\ 
  (1,1,1) & 1000 & M & 0.863(0.0037) & 0.943(0.00521) & 13.7(0.611) & 27.2(1.22) & 13.5(0.0581) & 13.6(0.061) \\ 
  (1,1,1) & 1000 & Y & 0.00272(0.00243) & 0.0014(0.00259) & 5.92(0.265) & 6.7(0.3) & 5.74(0.0115) & 5.76(0.014) \\ 
   \hline

\end{longtable}
\end{landscape}

\begin{landscape}
\centering
\begin{longtable}{|c|c|c|c|c|c|c|c|c|c|}
\caption{NIDE estimand for Data Generating Process A}
\label{nide_dgpA}
\endfirsthead
\endhead
\hline
Beta & N & Spec & Bias\_G & Bias\_TTS & Var\_G & Var\_TTS & Theory\_Var\_G & Boot\_Var\_G \\ 
  \hline
(0,0,0) & 100 & XYM & -0.000762(0.00069) & -0.000561(0.00123) & 0.0476(0.00213) & 0.151(0.00675) & 0.0974(0.00342) & 0.154(0.154) \\ 
  (0,0,0) & 100 & XY & 0.00211(0.000644) & 0.00228(0.00151) & 0.0414(0.00185) & 0.229(0.0103) & 0.0731(0.00303) & 0.112(0.112) \\ 
  (0,0,0) & 100 & XM & 0.00241(0.00128) & 0.0055(0.0022) & 0.163(0.0073) & 0.486(0.0217) & 0.29(0.0112) & 0.445(0.445) \\ 
  (0,0,0) & 100 & M & 0.000992(0.00121) & 0.00284(0.00223) & 0.146(0.00651) & 0.497(0.0223) & 0.271(0.0101) & 0.413(0.413) \\ 
  (0,0,0) & 100 & Y & 0.00209(0.00181) & 0.00264(0.00228) & 0.328(0.0147) & 0.518(0.0232) & 0.311(0.0075) & 0.36(0.36) \\ 
  (0,0,0) & 500 & XYM & -2.61e-05(0.000136) & -0.000147(0.000248) & 0.0093(0.000416) & 0.0308(0.00138) & 0.019(0.000606) & 0.029(0.029) \\ 
  (0,0,0) & 500 & XY & 9.12e-06(0.000105) & -6.06e-06(0.000298) & 0.00556(0.000249) & 0.0445(0.00199) & 0.0125(0.000427) & 0.0192(0.0192) \\ 
  (0,0,0) & 500 & XM & -0.000496(0.000261) & -0.000989(0.000548) & 0.0341(0.00153) & 0.15(0.00673) & 0.0621(0.00217) & 0.0921(0.0921) \\ 
  (0,0,0) & 500 & M & 0.000276(0.000243) & 0.000609(0.000412) & 0.0295(0.00132) & 0.0847(0.00379) & 0.0536(0.00184) & 0.0796(0.0796) \\ 
  (0,0,0) & 500 & Y & -0.00153(0.000734) & -0.00077(0.000933) & 0.27(0.0121) & 0.436(0.0195) & 0.27(0.00291) & 0.279(0.279) \\ 
  (0,0,0) & 1000 & XYM & 2.98e-05(6.83e-05) & -0.000152(0.000112) & 0.00466(0.000209) & 0.0125(0.000561) & 0.00954(0.000296) & 0.0145(0.0145) \\ 
  (0,0,0) & 1000 & XY & -3.7e-05(5.58e-05) & -1.95e-05(0.000161) & 0.00312(0.00014) & 0.0259(0.00116) & 0.00646(0.000211) & 0.00974(0.00974) \\ 
  (0,0,0) & 1000 & XM & -2.77e-05(0.000126) & 5.37e-05(0.000268) & 0.0158(0.000706) & 0.0716(0.0032) & 0.0292(0.000954) & 0.0437(0.0437) \\ 
  (0,0,0) & 1000 & M & 4.46e-05(0.000115) & 4.94e-05(0.000194) & 0.0133(0.000594) & 0.0375(0.00168) & 0.0267(0.000878) & 0.0399(0.0399) \\ 
  (0,0,0) & 1000 & Y & -0.00055(0.000508) & -0.000439(0.000653) & 0.258(0.0115) & 0.426(0.0191) & 0.266(0.00217) & 0.269(0.269) \\ 
  (1,1,1) & 100 & XYM & 0.003(0.00743) & 0.011(0.0103) & 5.52(0.247) & 10.7(0.478) & 5.88(0.0507) & 6.16(6.16) \\ 
  (1,1,1) & 100 & XY & -0.000413(0.0103) & -0.00162(0.0178) & 10.5(0.472) & 31.7(1.42) & 9.58(0.109) & 9.8(9.8) \\ 
  (1,1,1) & 100 & XM & 0.00459(0.0093) & 0.0116(0.0145) & 8.65(0.387) & 21(0.942) & 8.03(0.0955) & 8.51(8.51) \\ 
  (1,1,1) & 100 & M & -0.00994(0.00879) & 0.0058(0.0134) & 7.73(0.346) & 17.9(0.802) & 7.39(0.0889) & 7.91(7.91) \\ 
  (1,1,1) & 100 & Y & 0.831(0.0111) & 0.921(0.0124) & 12.4(0.553) & 15.5(0.693) & 11.3(0.111) & 11.7(11.7) \\ 
  (1,1,1) & 500 & XYM & 0.000102(0.00352) & -0.00223(0.0044) & 6.19(0.277) & 9.67(0.433) & 5.84(0.0236) & 5.9(5.9) \\ 
  (1,1,1) & 500 & XY & -0.00721(0.00441) & -0.0119(0.00778) & 9.73(0.435) & 30.3(1.35) & 9.88(0.0587) & 9.9(9.9) \\ 
  (1,1,1) & 500 & XM & 0.00657(0.00394) & 0.00122(0.00917) & 7.77(0.348) & 42.1(1.88) & 7.92(0.0416) & 8.03(8.03) \\ 
  (1,1,1) & 500 & M & -0.00231(0.00376) & -0.00462(0.00599) & 7.06(0.316) & 17.9(0.803) & 7.42(0.0403) & 7.58(7.58) \\ 
  (1,1,1) & 500 & Y & 0.855(0.00516) & 0.937(0.0055) & 13.3(0.596) & 15.1(0.677) & 11.8(0.0618) & 11.8(11.8) \\ 
  (1,1,1) & 1000 & XYM & -0.00296(0.00241) & -0.00196(0.00314) & 5.83(0.261) & 9.83(0.44) & 5.84(0.0181) & 5.87(5.87) \\ 
  (1,1,1) & 1000 & XY & 0.00151(0.00319) & -0.00187(0.00545) & 10.2(0.454) & 29.7(1.33) & 9.99(0.0444) & 10(10) \\ 
  (1,1,1) & 1000 & XM & 0.00169(0.00286) & -0.00207(0.00693) & 8.18(0.366) & 48.1(2.15) & 7.89(0.0308) & 7.93(7.93) \\ 
  (1,1,1) & 1000 & M & -0.000208(0.0028) & 0.00131(0.00458) & 7.81(0.35) & 21(0.938) & 7.45(0.0309) & 7.6(7.6) \\ 
  (1,1,1) & 1000 & Y & 0.865(0.00339) & 0.946(0.00361) & 11.5(0.514) & 13.1(0.584) & 11.8(0.0447) & 11.8(11.8) \\ 
   \hline

\end{longtable}
\end{landscape}

\begin{landscape}
\centering
\begin{longtable}{|c|c|c|c|c|c|c|c|c|c|}
\caption{NDE estimand for Data Generating Process B}
\label{nde_dgpB}
\endfirsthead
\endhead
\hline
Beta & N & Spec & Bias\_G & Bias\_TTS & Var\_G & Var\_TTS & Theory\_Var\_G & Boot\_Var\_G \\ 
  \hline
(0,0,0) & 100 & XYM & -0.00368(0.00703) & -0.00329(0.00757) & 4.95(0.221) & 5.73(0.256) & 4.8(0.0329) & 5.02(0.0354) \\ 
  (0,0,0) & 100 & XY & -0.0067(0.00723) & -0.00355(0.00782) & 5.22(0.234) & 6.11(0.273) & 4.8(0.0342) & 5(0.036) \\ 
  (0,0,0) & 100 & XM & 0.00277(0.00967) & 0.0136(0.0149) & 9.35(0.419) & 22.2(0.993) & 9.1(0.0968) & 9.44(0.097) \\ 
  (0,0,0) & 100 & M & 0.859(0.0106) & 0.946(0.0112) & 11.2(0.499) & 12.5(0.558) & 10.1(0.0966) & 10.5(0.0994) \\ 
  (0,0,0) & 100 & Y & 0.014(0.00713) & 0.0153(0.00731) & 5.09(0.228) & 5.34(0.239) & 4.7(0.029) & 4.87(0.0308) \\ 
  (0,0,0) & 500 & XYM & 0.000774(0.00305) & -0.000145(0.00323) & 4.65(0.208) & 5.23(0.234) & 4.81(0.0137) & 4.85(0.0154) \\ 
  (0,0,0) & 500 & XY & 0.00594(0.00324) & 0.00598(0.00343) & 5.24(0.234) & 5.87(0.263) & 4.83(0.0136) & 4.85(0.0155) \\ 
  (0,0,0) & 500 & XM & -0.00153(0.00435) & -0.00482(0.00808) & 9.45(0.423) & 32.7(1.46) & 9.63(0.0557) & 9.69(0.0565) \\ 
  (0,0,0) & 500 & M & 0.864(0.00459) & 0.947(0.00475) & 10.5(0.472) & 11.3(0.504) & 10.6(0.0526) & 10.7(0.0541) \\ 
  (0,0,0) & 500 & Y & -0.00428(0.00308) & -0.00494(0.00313) & 4.73(0.212) & 4.89(0.219) & 4.75(0.012) & 4.78(0.0138) \\ 
  (0,0,0) & 1000 & XYM & 0.000397(0.00226) & 0.00151(0.00238) & 5.12(0.229) & 5.65(0.253) & 4.83(0.00957) & 4.86(0.0116) \\ 
  (0,0,0) & 1000 & XY & -0.000543(0.00224) & -0.00172(0.00232) & 5(0.224) & 5.38(0.241) & 4.82(0.00953) & 4.83(0.012) \\ 
  (0,0,0) & 1000 & XM & 0.00162(0.00325) & 0.00184(0.00547) & 10.6(0.473) & 30(1.34) & 9.67(0.0395) & 9.7(0.0425) \\ 
  (0,0,0) & 1000 & M & 0.868(0.00323) & 0.948(0.00333) & 10.5(0.468) & 11.1(0.498) & 10.6(0.0381) & 10.7(0.0415) \\ 
  (0,0,0) & 1000 & Y & 0.00221(0.00214) & 0.00212(0.00218) & 4.58(0.205) & 4.73(0.212) & 4.76(0.0087) & 4.78(0.011) \\ 
  (1,1,1) & 100 & XYM & 0.012(0.00748) & 0.0165(0.0104) & 5.59(0.25) & 10.7(0.481) & 5.48(0.0399) & 5.76(0.0431) \\ 
  (1,1,1) & 100 & XY & 0.0157(0.00721) & 0.0209(0.00754) & 5.2(0.233) & 5.68(0.254) & 5.07(0.0358) & 5.29(0.0377) \\ 
  (1,1,1) & 100 & XM & 0.000393(0.0108) & -0.0156(0.0217) & 11.7(0.525) & 47.3(2.12) & 11(0.127) & 11.5(0.128) \\ 
  (1,1,1) & 100 & M & 0.849(0.0109) & 0.917(0.0166) & 11.9(0.532) & 27.6(1.24) & 11.7(0.133) & 12.2(0.136) \\ 
  (1,1,1) & 100 & Y & 0.00425(0.00763) & 0.0164(0.00931) & 5.82(0.261) & 8.66(0.388) & 5.47(0.0368) & 5.71(0.0393) \\ 
  (1,1,1) & 500 & XYM & 0.000856(0.00326) & -0.00715(0.0112) & 5.33(0.238) & 63.1(2.82) & 5.4(0.0175) & 5.45(0.0193) \\ 
  (1,1,1) & 500 & XY & -0.00366(0.00313) & -0.00349(0.00337) & 4.91(0.22) & 5.67(0.254) & 5.12(0.0148) & 5.16(0.0159) \\ 
  (1,1,1) & 500 & XM & 0.00671(0.0049) & -0.0199(0.0274) & 12(0.537) & 374(16.7) & 11.3(0.0617) & 11.3(0.0635) \\ 
  (1,1,1) & 500 & M & 0.862(0.00503) & 1.19(0.24) & 12.7(0.567) & 28800(1290) & 12.4(0.0697) & 12.5(0.0707) \\ 
  (1,1,1) & 500 & Y & -0.000415(0.00329) & -0.00134(0.00975) & 5.41(0.242) & 47.5(2.13) & 5.5(0.0164) & 5.53(0.018) \\ 
  (1,1,1) & 1000 & XYM & 0.00236(0.00228) & 0.0089(0.0455) & 5.21(0.233) & 2070(92.6) & 5.43(0.0119) & 5.46(0.0144) \\ 
  (1,1,1) & 1000 & XY & -0.000313(0.00221) & -0.00102(0.00234) & 4.87(0.218) & 5.48(0.245) & 5.1(0.0106) & 5.12(0.0129) \\ 
  (1,1,1) & 1000 & XM & 0.00758(0.00329) & 0.113(0.0773) & 10.8(0.483) & 5980(268) & 11.4(0.0459) & 11.5(0.048) \\ 
  (1,1,1) & 1000 & M & 0.864(0.00339) & 0.87(0.0548) & 11.5(0.513) & 3010(135) & 12.3(0.0469) & 12.3(0.0508) \\ 
  (1,1,1) & 1000 & Y & -0.000468(0.00235) & -0.199(0.221) & 5.51(0.247) & 49000(2190) & 5.5(0.0108) & 5.52(0.0133) \\ 
   \hline

\end{longtable}
\end{landscape}

\begin{landscape}
\centering
\begin{longtable}{|c|c|c|c|c|c|c|c|c|c|}
\caption{NIDE estimand for Data Generating Process B}
\label{nide_dgpB}
\endfirsthead
\endhead
\hline
Beta & N & Spec & Bias\_G & Bias\_TTS & Var\_G & Var\_TTS & Theory\_Var\_G & Boot\_Var\_G \\ 
  \hline
(0,0,0) & 100 & XYM & 0.000217(0.000709) & 0.00168(0.00121) & 0.0503(0.00225) & 0.147(0.00659) & 0.0984(0.00356) & 0.155(0.155) \\ 
  (0,0,0) & 100 & XY & 0.000536(0.000679) & 0.00193(0.00154) & 0.0461(0.00206) & 0.237(0.0106) & 0.0742(0.00289) & 0.117(0.117) \\ 
  (0,0,0) & 100 & XM & -0.000557(0.00124) & -0.00262(0.00261) & 0.154(0.00688) & 0.679(0.0304) & 0.285(0.0112) & 0.441(0.441) \\ 
  (0,0,0) & 100 & M & 0.000394(0.00114) & 0.0045(0.00231) & 0.13(0.00582) & 0.535(0.0239) & 0.242(0.0106) & 0.381(0.381) \\ 
  (0,0,0) & 100 & Y & -0.000822(0.0016) & -0.0018(0.00223) & 0.258(0.0115) & 0.498(0.0223) & 0.28(0.00735) & 0.332(0.332) \\ 
  (0,0,0) & 500 & XYM & 8.56e-05(0.000149) & 0.000237(0.000246) & 0.0111(0.000495) & 0.0302(0.00135) & 0.0196(0.000666) & 0.0296(0.0296) \\ 
  (0,0,0) & 500 & XY & 1.89e-05(0.000119) & 0.000217(0.000285) & 0.00702(0.000314) & 0.0407(0.00182) & 0.0141(0.000477) & 0.0213(0.0213) \\ 
  (0,0,0) & 500 & XM & 3.9e-05(0.000242) & -0.000221(0.000528) & 0.0293(0.00131) & 0.139(0.00624) & 0.0571(0.00189) & 0.0856(0.0856) \\ 
  (0,0,0) & 500 & M & 1.1e-05(0.000228) & 0.000383(0.000398) & 0.026(0.00116) & 0.0794(0.00355) & 0.0517(0.00181) & 0.0779(0.0779) \\ 
  (0,0,0) & 500 & Y & 0.000357(0.000675) & 0.000558(0.000943) & 0.228(0.0102) & 0.444(0.0199) & 0.222(0.00265) & 0.23(0.23) \\ 
  (0,0,0) & 1000 & XYM & 1.94e-05(7.09e-05) & -0.000156(0.000121) & 0.00502(0.000225) & 0.0146(0.000652) & 0.00985(0.000324) & 0.0148(0.0148) \\ 
  (0,0,0) & 1000 & XY & 4.19e-05(6.36e-05) & 7.49e-05(0.000152) & 0.00405(0.000181) & 0.023(0.00103) & 0.00729(0.000251) & 0.0108(0.0108) \\ 
  (0,0,0) & 1000 & XM & -0.000164(0.000122) & -4.87e-05(0.000261) & 0.0149(0.000665) & 0.0681(0.00305) & 0.0293(0.000969) & 0.044(0.044) \\ 
  (0,0,0) & 1000 & M & -0.000187(0.000119) & -0.000146(0.000199) & 0.0142(0.000636) & 0.0396(0.00177) & 0.0258(0.000902) & 0.0388(0.0388) \\ 
  (0,0,0) & 1000 & Y & -0.000691(0.000461) & -0.000609(0.000637) & 0.213(0.00953) & 0.406(0.0182) & 0.219(0.00194) & 0.223(0.223) \\ 
  (1,1,1) & 100 & XYM & -0.0104(0.00965) & -0.017(0.0121) & 9.3(0.416) & 14.6(0.652) & 8.7(0.11) & 9.09(9.09) \\ 
  (1,1,1) & 100 & XY & -0.0215(0.0116) & -0.0301(0.0175) & 13.4(0.6) & 30.5(1.36) & 12.6(0.145) & 12.9(12.9) \\ 
  (1,1,1) & 100 & XM & 0.00847(0.00995) & 0.00372(0.0188) & 9.89(0.443) & 35.2(1.57) & 9.97(0.128) & 10.6(10.6) \\ 
  (1,1,1) & 100 & M & -0.0017(0.00999) & 0.0244(0.0162) & 9.97(0.446) & 26.2(1.17) & 9.42(0.129) & 10(10) \\ 
  (1,1,1) & 100 & Y & 0.834(0.0119) & 0.906(0.0135) & 14.1(0.632) & 18.1(0.811) & 13.8(0.141) & 14.3(14.3) \\ 
  (1,1,1) & 500 & XYM & -0.0118(0.00416) & -0.00184(0.0116) & 8.65(0.387) & 67.8(3.04) & 8.61(0.0449) & 8.68(8.68) \\ 
  (1,1,1) & 500 & XY & 0.0018(0.00523) & -0.00395(0.00823) & 13.7(0.611) & 33.9(1.52) & 13.1(0.0755) & 13.2(13.2) \\ 
  (1,1,1) & 500 & XM & 0.000698(0.00445) & 0.0333(0.0266) & 9.89(0.443) & 353(15.8) & 9.88(0.0565) & 9.99(9.99) \\ 
  (1,1,1) & 500 & M & 0.00828(0.00433) & -0.238(0.24) & 9.37(0.419) & 28800(1290) & 9.39(0.0647) & 9.65(9.65) \\ 
  (1,1,1) & 500 & Y & 0.855(0.00529) & 0.938(0.0106) & 14(0.626) & 55.8(2.5) & 14.5(0.0749) & 14.6(14.6) \\ 
  (1,1,1) & 1000 & XYM & 0.00217(0.00298) & -0.00639(0.0457) & 8.9(0.398) & 2090(93.3) & 8.71(0.0361) & 8.75(8.75) \\ 
  (1,1,1) & 1000 & XY & -0.00308(0.00367) & -0.0062(0.00601) & 13.5(0.603) & 36.1(1.62) & 13.2(0.0548) & 13.2(13.2) \\ 
  (1,1,1) & 1000 & XM & 0.00129(0.00319) & -0.0985(0.0772) & 10.2(0.456) & 5960(267) & 9.84(0.0422) & 9.92(9.92) \\ 
  (1,1,1) & 1000 & M & -0.00305(0.00302) & 0.0711(0.055) & 9.12(0.408) & 3020(135) & 9.39(0.0538) & 9.55(9.55) \\ 
  (1,1,1) & 1000 & Y & 0.862(0.00374) & 1.14(0.221) & 14(0.624) & 49000(2190) & 14.7(0.0584) & 14.7(14.7) \\ 
   \hline

\end{longtable}
\end{landscape}

\begin{landscape}
\centering
\begin{longtable}{|c|c|c|c|c|c|c|c|c|c|}
\caption{NDE estimand for Data Generating Process C}
\label{nde_dgpC}
\endfirsthead
\endhead
\hline
Beta & N & Spec & Bias\_G & Bias\_TTS & Var\_G & Var\_TTS & Theory\_Var\_G & Boot\_Var\_G \\ 
  \hline
(0,0,0) & 100 & XYM & -0.00301(0.00656) & 0.00107(0.00705) & 4.31(0.193) & 4.97(0.222) & 4.77(0.0313) & 6.54(0.248) \\ 
  (0,0,0) & 100 & XY & 0.00514(0.00733) & 0.00348(0.00803) & 5.37(0.24) & 6.45(0.288) & 4.79(0.0321) & 7.62(0.303) \\ 
  (0,0,0) & 100 & XM & 0.00579(0.0101) & 0.0267(0.0184) & 10.2(0.455) & 34(1.52) & 9.3(0.122) & 12.3(0.503) \\ 
  (0,0,0) & 100 & M & 0.836(0.0105) & 0.92(0.0111) & 11(0.494) & 12.3(0.552) & 9.96(0.0995) & 80.9(1.9) \\ 
  (0,0,0) & 100 & Y & 0.00444(0.00685) & 0.00214(0.00691) & 4.69(0.21) & 4.78(0.214) & 4.59(0.0306) & 7.85(0.311) \\ 
  (0,0,0) & 500 & XYM & 0.000106(0.00301) & 0.00157(0.00314) & 4.52(0.202) & 4.94(0.221) & 4.83(0.0136) & 6.86(0.264) \\ 
  (0,0,0) & 500 & XY & 0.000599(0.00315) & 0.0014(0.00335) & 4.98(0.223) & 5.62(0.252) & 4.83(0.0139) & 6.85(0.267) \\ 
  (0,0,0) & 500 & XM & -0.00674(0.00455) & -0.00605(0.00887) & 10.4(0.464) & 39.3(1.76) & 9.64(0.0593) & 12.3(0.488) \\ 
  (0,0,0) & 500 & M & 0.862(0.00472) & 0.944(0.0048) & 11.1(0.499) & 11.5(0.515) & 10.5(0.0534) & 378(4.19) \\ 
  (0,0,0) & 500 & Y & 0.000134(0.00313) & 0.000521(0.00314) & 4.9(0.219) & 4.93(0.221) & 4.61(0.0121) & 14.2(0.497) \\ 
  (0,0,0) & 1000 & XYM & -0.00227(0.00225) & -0.00227(0.00238) & 5.07(0.227) & 5.67(0.254) & 4.83(0.00983) & 7.16(0.288) \\ 
  (0,0,0) & 1000 & XY & -0.00111(0.00221) & -0.000989(0.0023) & 4.88(0.218) & 5.28(0.236) & 4.84(0.0101) & 7.07(0.258) \\ 
  (0,0,0) & 1000 & XM & -0.00127(0.00302) & 0.000813(0.00527) & 9.12(0.408) & 27.8(1.24) & 9.64(0.0377) & 11.7(0.472) \\ 
  (0,0,0) & 1000 & M & 0.863(0.00337) & 0.942(0.0034) & 11.4(0.509) & 11.6(0.518) & 10.6(0.0371) & 742(5.98) \\ 
  (0,0,0) & 1000 & Y & -0.000594(0.00219) & -0.00018(0.00222) & 4.8(0.215) & 4.91(0.22) & 4.64(0.00849) & 21.4(0.633) \\ 
  (1,1,1) & 100 & XYM & 0.00449(0.00736) & 0.00319(0.00818) & 5.42(0.242) & 6.69(0.299) & 5.11(0.0375) & 70.7(1.34) \\ 
  (1,1,1) & 100 & XY & -0.00356(0.00714) & 0.00596(0.00833) & 5.1(0.228) & 6.95(0.311) & 4.97(0.0351) & 73.6(1.33) \\ 
  (1,1,1) & 100 & XM & 0.0209(0.0103) & -0.00588(0.0212) & 10.6(0.474) & 45.1(2.02) & 9.77(0.121) & 79.3(1.83) \\ 
  (1,1,1) & 100 & M & 0.888(0.0109) & 0.924(0.0115) & 11.9(0.534) & 13.3(0.595) & 10.9(0.108) & 298(3.81) \\ 
  (1,1,1) & 100 & Y & -0.00842(0.00693) & -0.00814(0.00754) & 4.81(0.215) & 5.69(0.255) & 5.03(0.0372) & 57.5(1.1) \\ 
  (1,1,1) & 500 & XYM & -0.000439(0.00312) & -0.000264(0.00336) & 4.87(0.218) & 5.64(0.252) & 5.06(0.0155) & 323(2.79) \\ 
  (1,1,1) & 500 & XY & -0.00397(0.0032) & -0.00326(0.00347) & 5.11(0.229) & 6.03(0.27) & 5.04(0.0154) & 336(2.86) \\ 
  (1,1,1) & 500 & XM & 0.0303(0.00452) & -0.00829(0.0084) & 10.2(0.457) & 35.3(1.58) & 10.2(0.0614) & 352(3.92) \\ 
  (1,1,1) & 500 & M & 0.901(0.00479) & 0.935(0.00495) & 11.5(0.513) & 12.3(0.548) & 11.2(0.0545) & 1460(8.46) \\ 
  (1,1,1) & 500 & Y & 0.00265(0.00313) & 0.00292(0.00328) & 4.91(0.219) & 5.38(0.241) & 5.02(0.0151) & 267(2.49) \\ 
  (1,1,1) & 1000 & XYM & -0.00664(0.0022) & -0.00507(0.00238) & 4.83(0.216) & 5.65(0.253) & 5.07(0.0111) & 629(3.77) \\ 
  (1,1,1) & 1000 & XY & -0.00122(0.00242) & -0.000419(0.00257) & 5.86(0.262) & 6.6(0.296) & 5.04(0.0108) & 675(4.3) \\ 
  (1,1,1) & 1000 & XM & 0.029(0.00318) & -0.0139(0.00569) & 10.1(0.453) & 32.4(1.45) & 10.4(0.0411) & 691(5.43) \\ 
  (1,1,1) & 1000 & M & 0.918(0.00339) & 0.951(0.00349) & 11.5(0.515) & 12.2(0.546) & 11.5(0.0405) & 2970(12.1) \\ 
  (1,1,1) & 1000 & Y & 0.000258(0.00229) & -0.000654(0.00239) & 5.24(0.234) & 5.73(0.257) & 5.05(0.0111) & 523(3.62) \\ 
   \hline

\end{longtable}
\end{landscape}

\begin{landscape}
\centering
\begin{longtable}{|c|c|c|c|c|c|c|c|c|c|}
\caption{NIDE estimand for Data Generating Process C}
\label{nide_dgpC}
\endfirsthead
\endhead
\hline
Beta & N & Spec & Bias\_G & Bias\_TTS & Var\_G & Var\_TTS & Theory\_Var\_G & Boot\_Var\_G \\ 
  \hline
(0,0,0) & 100 & XYM & -0.000386(0.000758) & 1.94e-05(0.0011) & 0.0574(0.00257) & 0.12(0.00537) & 0.0994(0.273) & 9.98(9.98) \\ 
  (0,0,0) & 100 & XY & 0.000943(0.000678) & 0.00172(0.00138) & 0.0459(0.00206) & 0.19(0.00849) & 0.0945(0.21) & 9.26(9.26) \\ 
  (0,0,0) & 100 & XM & 0.00118(0.00118) & 0.00274(0.00183) & 0.14(0.00625) & 0.336(0.015) & 0.241(0.596) & 22.5(22.5) \\ 
  (0,0,0) & 100 & M & -0.00348(0.00116) & -0.00133(0.00158) & 0.135(0.00602) & 0.251(0.0112) & 0.224(0.58) & 21.9(21.9) \\ 
  (0,0,0) & 100 & Y & -0.00104(0.00105) & -0.000516(0.00147) & 0.111(0.00495) & 0.216(0.00967) & 0.159(0.179) & 8.59(8.59) \\ 
  (0,0,0) & 500 & XYM & 0.000136(0.000144) & -0.000222(0.000225) & 0.0103(0.000461) & 0.0252(0.00113) & 0.0198(0.223) & 9.59(9.59) \\ 
  (0,0,0) & 500 & XY & -0.000105(0.000137) & -7.15e-05(0.00026) & 0.00934(0.000418) & 0.0339(0.00152) & 0.0192(0.221) & 9.3(9.3) \\ 
  (0,0,0) & 500 & XM & 0.000125(0.000217) & 0.00019(0.00033) & 0.0234(0.00105) & 0.0543(0.00243) & 0.0461(0.506) & 21.8(21.8) \\ 
  (0,0,0) & 500 & M & -0.000259(0.000191) & 0.000187(0.000271) & 0.0183(0.000817) & 0.0369(0.00165) & 0.0392(0.467) & 20.2(20.2) \\ 
  (0,0,0) & 500 & Y & 0.000414(0.000398) & 0.000288(0.000525) & 0.079(0.00354) & 0.138(0.00616) & 0.0888(0.16) & 8.2(8.2) \\ 
  (0,0,0) & 1000 & XYM & -6.41e-05(7.27e-05) & 5.69e-05(0.000113) & 0.00529(0.000237) & 0.0128(0.000571) & 0.0101(0.202) & 9.45(9.45) \\ 
  (0,0,0) & 1000 & XY & 3.72e-05(6.75e-05) & 2.97e-05(0.000135) & 0.00456(0.000204) & 0.0182(0.000813) & 0.0099(0.193) & 9.01(9.01) \\ 
  (0,0,0) & 1000 & XM & -1.45e-05(0.000103) & -9.14e-05(0.000177) & 0.0107(0.000477) & 0.0315(0.00141) & 0.0227(0.419) & 20.6(20.6) \\ 
  (0,0,0) & 1000 & M & -0.000306(0.000115) & 9.28e-05(0.000152) & 0.0133(0.000595) & 0.0231(0.00103) & 0.0224(0.506) & 21.3(21.3) \\ 
  (0,0,0) & 1000 & Y & 6.49e-05(0.00027) & -0.00032(0.000358) & 0.0727(0.00325) & 0.128(0.00572) & 0.0769(0.162) & 8.28(8.28) \\ 
  (1,1,1) & 100 & XYM & 0.0155(0.00379) & 0.00598(0.0045) & 1.44(0.0643) & 2.02(0.0905) & 1.39(1.13) & 73.4(73.4) \\ 
  (1,1,1) & 100 & XY & 0.0132(0.00346) & -0.00869(0.00528) & 1.2(0.0535) & 2.78(0.125) & 1.23(1.21) & 78.1(78.1) \\ 
  (1,1,1) & 100 & XM & -0.033(0.00366) & -0.00466(0.00569) & 1.34(0.0599) & 3.24(0.145) & 1.43(1.29) & 62.1(62.1) \\ 
  (1,1,1) & 100 & M & -0.0495(0.0036) & -0.00195(0.00502) & 1.29(0.0579) & 2.52(0.113) & 1.35(1.19) & 56.9(56.9) \\ 
  (1,1,1) & 100 & Y & 0.112(0.00395) & 0.11(0.00469) & 1.56(0.0699) & 2.2(0.0985) & 1.57(1) & 63.5(63.5) \\ 
  (1,1,1) & 500 & XYM & 0.00586(0.00163) & -0.00301(0.00181) & 1.32(0.0591) & 1.65(0.0736) & 1.26(2.43) & 322(322) \\ 
  (1,1,1) & 500 & XY & 0.0154(0.00148) & 0.00137(0.00175) & 1.1(0.0492) & 1.53(0.0683) & 1.14(2.49) & 343(343) \\ 
  (1,1,1) & 500 & XM & -0.0269(0.00157) & 0.000325(0.00247) & 1.24(0.0553) & 3.05(0.137) & 1.26(2.57) & 234(234) \\ 
  (1,1,1) & 500 & M & -0.0453(0.00149) & -0.00169(0.00193) & 1.12(0.0499) & 1.86(0.0833) & 1.13(2.63) & 206(206) \\ 
  (1,1,1) & 500 & Y & 0.111(0.00167) & 0.107(0.00192) & 1.4(0.0626) & 1.84(0.0823) & 1.44(2.08) & 268(268) \\ 
  (1,1,1) & 1000 & XYM & 0.00812(0.00112) & -0.00125(0.00122) & 1.25(0.0558) & 1.49(0.0669) & 1.26(3.39) & 644(644) \\ 
  (1,1,1) & 1000 & XY & 0.0115(0.00106) & -0.00175(0.00123) & 1.13(0.0505) & 1.51(0.0677) & 1.12(3.57) & 675(675) \\ 
  (1,1,1) & 1000 & XM & -0.0276(0.00111) & -0.000538(0.00162) & 1.23(0.0549) & 2.64(0.118) & 1.24(3.78) & 451(451) \\ 
  (1,1,1) & 1000 & M & -0.0464(0.00102) & -0.0035(0.00134) & 1.05(0.0469) & 1.81(0.0809) & 1.09(3.57) & 385(385) \\ 
  (1,1,1) & 1000 & Y & 0.113(0.0012) & 0.109(0.00141) & 1.45(0.0647) & 1.98(0.0888) & 1.44(2.92) & 529(529) \\ 
   \hline

\end{longtable}
\end{landscape}

\end{document}